\documentclass[12pt]{article}

\usepackage{latexsym}
\usepackage{amsfonts}
\usepackage{amsmath}

\title{On Conserved Quantities at Spatial Infinity\thanks
{This research has been presented as a thesis
to the Department of Physics, University of Chicago, in
partial fulfillment of the requirements of the Ph.D. degree.
}}

\author{Shyan-Ming Perng\thanks{E-mail address: {\tt sperng@rainbow.uchicago.edu}}\\
\\
 {\em Enrico Fermi Institute, University of Chicago}\\
 {\em 5640 S. Ellis Ave. Chicago, Illinois, 60637 }
}

\renewcommand{\thesection}{\Roman{section}.}
\renewcommand{\thesubsection}{\Alph{subsection}.}

\newcommand{\be}{\begin{equation}}
\newcommand{\ee}{\end{equation}}
\newcommand{\bea}{\begin{eqnarray}}
\newcommand{\eea}{\end{eqnarray}}
\newcommand{\bi}{\begin{itemize}}
\newcommand{\ei}{\end{itemize}}

\newcommand{\met}[2]{\mbox{$\stackrel{#2}{#1}$}}
\newcommand{\metbar}[2]{\mbox{$\stackrel{#2}{{#1}}$}}
\newcommand{\ldb}{\mbox{$\stackrel{1}{{\lambda}}$}}

\newtheorem{theorem}{Theorem}

\newtheorem{definition}{Definition}

\begin{document}

%
%
\newcommand{\hspca}{\hspace{-0.50cm}}
\newcommand{\hspcb}{\hspace{-0.36cm}}

%
%
\baselineskip=24pt

\maketitle

\begin{abstract}
There is a well-known short list of asymptotic conserved quantities
for a physical system at spatial infinity. 
We search for new ones.
This is carried out
within the asymptotic framework of Ashtekar and Romano, 
in which spatial infinity is represented as a smooth boundary of space-time. 
We first introduce, for physical fields on space-time,
a characterization of their asymptotic behavior
as certain fields on this boundary. 
Conserved quantities at spatial infinity, in turn, are
constructed from these fields. 
We find,
in Minkowski space-time, that each of a Klein-Gordon field,
a Maxwell field, and a linearized gravitational field
yields an entire hierarchy of conserved quantities. 
Only certain quantities in this hierarchy survive into curved space-time.
\end{abstract}


\noindent Running title: Asymptotic Conserved Quantities

\newpage

\section{INTRODUCTION}

In the description of isolated systems in flat space-time, 
conserved quantities have often been found to be useful. Examples
of such conserved quantities include 
electric charge, energy-momentum, angular momentum, 
and, in certain circumstances, various multipole moments. 
These conserved quantities are usually expressed as surface integrals
in the limit as the surface approaches infinity.
In general relativity, by contrast,
the construction of such conserved quantities
is more complicated.
Not least of these complications is that ``infinity''
is so much more difficult to pin down in the presence of curvature.

The study of isolated systems in general relativity
was pioneered by Arnowitt, Deser and Misner$^1$
They defined asymptotic flatness of a space-time in 
terms of the existence of 
an initial-data set which, expressed in suitable
coordinates, has the initial data approach the flat
values at suitable rates.
Conserved quantities, such as energy-momentum and angular momentum,
were then be expressed as limits of certain 
surface integrals.

One unfortunate aspect of the approach 
of Arnowitt, Deser and Misner 
is that their asymptotic conditions 
are tied so closely to coordinates.
Their approach was subsequently geometrized and extended
by  Geroch$^2$
via a conformal 
completion by a single point ``at spatial infinity''.
Multipole moments for certain fields in flat space-time were
generalized to static asymptotically flat space-times
within this framework$^3$
%
%
An alternative geometrical framework,
which unifies spatial and null infinity
and is thus adapted to the relation between these two
asymptotic regimes,
was introduced by Ashtekar and Hansen$^4$
This framework involves a conformal completion of the
entire space-time,  null infinity becoming a null cone with spatial
infinity its vertex. 
This framework is used, for example, both to formulate and to prove 
the assertion that the ADM mass is the past limit of 
the future Bondi mass$^5$.

In both of the geometrical frameworks outlined above
spatial infinity is squeezed into a point, 
and there smoothness of the completed manifold fails. 
So, inevitably, one is forced to deal with 
complicated differentiable structures there. 
This circumstance is less satisfactory than that of null infinity,
which is formulated as a smooth boundary of space-time.
Early attempts to restore smoothness to spatial infinity
include those of Sommers$^6$
and Persides$^7$.
%
Beig and Schmidt$^{8,9}$,
using a coordinate-dependent treatment similar to
that of Bondi et. al$^9$,
obtained fields on the surface at spatial infinity order by order,
and noticed that these fields there satisfy hyperbolic equations.
This work culminates in that of Ashtekar and Romano$^{11}$,
who introduced a new geometrical framework for asymptotic flatness
in which spatial infinity was indeed expressed as
a smooth boundary of space-time. 
Their definition also provided a natural geometrical setting for the results
of Beig and Schmidt. 
Ashtekar and Romano's framework is somewhat of a 
hybrid, in that it involves both the conformal
and projective structure.
By their definition,
a space-time is asymptotically flat at spatial infinity provided one 
can attach to it a smooth boundary ${\cal H}$ and
introduce a smooth function $\Omega$
vanishing at ${\cal H}$ 
such that
the induced metric on and the normal 
to the constant-$\Omega$ surfaces are,
after rescaling by suitable powers of $\Omega$, 
smoothly extendible to ${\cal H}$. 
This new definition has proven  to be useful in 
the study of asymptotic properties of space-time at spatial infinity
since various physical fields turn out to be smooth there.

We return now to conserved quantities.
It is natural to ask: Do, in some sense, 
the well-known conserved quantities---energy-momentum,
angular momentum, electric charge---at spatial infinity
exhaust all conserved quantities that could possibly be defined
there?
To settle this question would clearly
provide insights into the asymptotic properties of 
the physical fields and of the space-time.
The framework introduced by Ashtekar and Romano is perfectly 
suited to
addressing this question. 
One has a simple, universal smooth structure at 
spatial infinity enabling one to
investigate fields at spatial infinity order by
order. 
The notion of a conserved quantity had
already been formulated by Ashtekar and Romano: 
Each conserved quantity is to be expressed as an integral over 
a 2-sphere section of spatial infinity 
where the value of the integral is independent of section. 
In particular,  the well-known conserved quantities are so
expressed. We seek others.

This paper is organized as follows. 
Section II contains the basic framework, which underlies
the rest of the paper. 
We first review briefly (a slight modification of)
the Ashtekar-Romano definition of asymptotic flatness.
We then formulate within this framework the asymptotic
structure of the physical fields.
In particular, we introduce the notion of a
conserved quantity, and give some familiar examples.
In section III, we consider the 
special but important case of fields in Minkowski
space-time. 
We construct all linear conserved quantities associated with
a Klein-Gordon field, 
with a Maxwell field, and 
with a linearized gravitational field 
and having a certain ``polynomial dependence'' on
asymptotic translations.
We then study
the symmetry properties 
and the ``gauge behavior'' 
(dependence on a certain freedom 
in the formulation of asymptotic structure)
of these quantities.
In section IV, we consider fields
in a curved, asymptotically flat space-time.
We first derive the 
equations, at spatial infinity,
satisfied by the asymptotic fields. 
We then show that---at least in the Klein-Gordon and Maxwell 
cases---certain of the conserved quantities found in section III
for Minkowski space-time can be generalized to these
curved space-times.
In section V, we discuss various related issues.
In particular, we formulate two conjectures. 
One asserts that a certain 
conserved quantity for linearized gravity in Minkowski
space-time can be generalized to curved space-time. 
The other asserts that we have here found all conserved
quantities in curved space-time for Klein-Gordon,
Maxwell and gravitational fields.

\section{PRELIMINARIES}

\subsection{Asymptotic Flatness}

Fix a space-time $(\tilde M,\tilde g_{ab})$.

\begin{definition}
By a {\em completion}\/ of $(\tilde M,\tilde g_{ab})$,
we mean (c.f. Ashtekar and Romano in Ref. 10): 
A manifold $M$
with boundary ${\cal H}$, a smooth function $\Omega$ defined
on $M$ vanishing on ${\cal H}$, 
and a diffeomorphism from $\tilde M$ to $M - {\cal H}$
(by means of which we identify $\tilde M$ with its image in $M$)
satisfying the following three classes of conditions: \\
$(1)\mbox{The combinations}\ (i)\nabla_a \Omega $,\  
$(ii)\Omega^{-4} \tilde g^{ab}\tilde\nabla_b\Omega \  (\equiv n^a )$,\ 
and $(iii)\Omega^2 [\tilde g_{ab} -
(\tilde g^{cd}\nabla_c\Omega\nabla_d\Omega )^{-1}
\nabla_a\Omega \nabla_b\Omega ]$\ $(\equiv q_{ab})$ 
admit smooth, nowhere-vanishing extensions to ${\cal H}$
such that $(iv)n^a\nabla_a\Omega (\equiv\lambda^{-2})|_{\cal H}=1$
and $(v){\cal L}_{n}[(n^m\nabla_m\Omega )^{-1}q_{ab}]|_{\cal H}=0$.
\\
$(2)({\cal H}, q_{ab}|_{\cal H})$ is a standard 
time-like hyperboloid,
i.e., ${\cal H}$ has topology $S^2\times \mathbb{R}$,
$q_{ab}|_{\cal H}\equiv \met{q}{0}{_{ab}}$ is of constant
positive curvature and is geodesically complete. 
\\
$(3)$The combinations $(i) n^kn^l\tilde G_{kl}$,
$(ii)\Omega^{-1}q_a{^k}n^l\tilde G_{kl}$, and
$(iii)\Omega^{-2}q_a{^k}q_b{^l}\tilde G_{kl}$, 
are smoothly extendible to ${\cal H}$,
where $\tilde G_{ab}$ is the Einstein tensor of $\tilde g_{ab}$.
\label{def-1}
\end{definition}
The boundary ${\cal H}$ represents spatial infinity.
Conditions $(1)$ describe the fall-off behavior of
the metric $\tilde g_{ab}$ and conditions (3) that of its second derivative.
Conditions (2) ensure, among other things, that we are dealing
with (all of) spatial infinity. 
There is some redundancy in the above conditions. 
Specifically, the constancy both of the left side of $(1)(iv)$
and of the curvature of $q_{ab}$ already follow from the other conditions.
In light of this, the choice of the constant ``1'' in
condition $(1)(iv)$ 
(which is equivalent to the demand that 
$\met{q}{0}{_{ab}}$ be the metric of a unit hyperboloid)
serves only to restrict the freedom of multiplying
$\Omega$ by a constant factor. 
Condition $(1)(v)$ is essentially
the condition, $\met{B}{0}{_{ab}} = 0$ (c.f. eqn.(\ref{weylebdef})),
introduced by Ashtekar and Hansen$^{4}$ 
in order to define angular momentum.
More precisely, when $\met{B}{0}{_{ab}} = 0$, 
condition $(1)(v)$ can always be achieved 
without affecting the other conditions
by choosing 
a suitable $\Omega$. 

Definition \ref{def-1}
is essentially the same as the definition, 
given by Ashtekar and Romano$^{11}$,
of what they call an asymptotically Min\-kow\-ski\-an space-time.
However, there are three differences.
First, our conditions on the Einstein tensor
are weaker than the corresponding condition, namely 
$\lim_{\Omega\rightarrow 0} \Omega^{-1}\tilde G_{ab}$
$=0$,
in their definition.
Their condition, 
expressed in the present language, is equivalent to
the smooth extendibility
to ${\cal H}$ of
$\Omega^{-2} n^kn^l\tilde G_{kl}$,
$\Omega^{-2}q_a{^k}n^l\tilde G_{kl}$, and
$\Omega^{-2}q_a{^k}q_b{^l}\tilde G_{kl}$. 
Indeed, our condition holds while theirs fails 
(for $n^kn^l\tilde G_{kl}|_{\cal H} \neq 0$)
in the Reissner-Nordstrom solution. 
Second, we impose condition $(1)(iv)$ which, as mentioned above, is
effectively a gauge restriction on the conformal factor $\Omega$,
a restriction that is absent in the definition of Ashtekar and Romano.
Finally, we impose condition $(1)(v)$,
which Ashtekar and Romano 
omit from the general definition of asymptotic Minkowskian
space-times, but subsequently impose 
for their discussion of angular momentum.

We give a few simple examples of completion. 
As a first example, 
let $(\tilde M,\tilde g_{ab})$ be Minkowski space-time,
and let $(t,r,\theta ,\phi )$ 
be ordinary spherical polar
coordinates. Set 
$\Omega =(r^2 - t^2)^{-1/2}$ and
$\tanh\chi = t /r$. 
Let $M$ be $\tilde M$ together
with the boundary ${\cal H}$ 
consisting of the
points labeled by $\Omega =0$ in the
(hyperbolic coordinate)
chart $(\Omega ,\chi ,\theta ,\phi )$, with differentiable
structure given by that chart. Then this $(M,\Omega )$
is a completion of Minkowski space-time.
%
As a second example, 
let $(\tilde M,\tilde g_{ab})$ be Reissner-Nordstrom solution,
and let $(t,r,\theta ,\phi )$ be the usual 
Schwarzschild-like coordinates therein.
Repeat the same construction as in Minkowski space-time
to obtain a manifold with boundary $(M,\Omega )$. 
Then this choice of $\Omega$
satisfies all conditions in definition \ref{def-1} except
condition $(1)(v)$.
Condition $(1)(v)$ in turn can be achieved, 
without violating other
conditions, by choosing a new conformal factor $\Omega '$ of
the form $\Omega ' = \Omega ( 1+\omega \Omega )$ with 
a suitable smooth function $\omega$.
In general, all stationary vacuum space-times asymptotically
flat by the usual definition$^{3}$ admit completions in the
present sense.$^{12}$

Two completions $(M,\Omega)$, $(M',\Omega')$
of $(\tilde M,\tilde g_{ab})$ are said to be equivalent if the
identity map of $\tilde M$
extends to a diffeomorphism from $M$ to $M'$.
It turns out that a space-time may admit inequivalent completions.$^{13}$
Minkowski space-time, for instance, has a four-parameter
family of inequivalent completions
related to each other by what are called
``logarithmic translations''.$^{14}$
Indeed, let $x^\mu$ be a usual Minkowskian coordinate system
in Minkowski space-time $\tilde M$,
and
$c^\mu$ any constant vector. Then
the hyperbolic coordinates associated with
${x'}^\mu$ given by
${x}^\mu = {x'}^\mu - c^\mu\log\Omega'$ 
yield a new completion of $\tilde M$ inequivalent
to that arising from $x^\mu$.
In this case we can single out the usual completion to
be the preferred one among this four-parameter family
since it is the only one in which all Killing fields
are smoothly extendible to the boundary at spatial 
infinity.
Similarly, any stationary asymptotically
flat space-time admits at least a one-parameter family
of inequivalent completions,
arising  from logarithmic time-translations.
There is also a sort of converse to this:
the existence of two inequivalent completions
related by such a logarithmic translation 
implies that the space-time admits 
an asymptotic translational Killing field---a vector field
$\tilde\xi^a$ with the properties that
$\Omega^{-1}\tilde\xi^a$ is smoothly extendible to,
and vanishes nowhere on, ${\cal H}$; and that
$\tilde\nabla_{(a}\tilde\xi_{b)}$ and all its derivatives
vanishes on ${\cal H}$.
In the spatial-infinity framework of Geroch and
Ashtekar-Hansen, it has been shown by Chrusciel$^{15}$
that these logarithmic translations are the only
kind of inequivalent completions that may arise.
We conjecture in the present framework:
Any two inequivalent completions are related by such a
logarithmic translation.
If this conjecture is true, then our work
will not be affected by the possible existence of
inequivalent completions. 
In what follows we will always fix 
a specific completion and only consider
completions smoothly related (i.e., equivalent) 
to the fixed one.

\subsection{Physical Fields and Their Remnants}

We now set up the framework for dealing with 
the asymptotic structure of physical fields.
Let $(\tilde M,\tilde g_{ab})$ be a space-time, 
with $(M,\Omega )$ a completion.
Let $\tilde v_{a_1\cdots a_m}$ be a smooth, covariant,
$m$-th rank tensor field 
on $\tilde M$, and consider 
the $2^m$ tensor fields that result from contracting each index
of $\tilde v$ with
either $\Omega^2n^a$ or $\Omega q^a{_b}$. 
We say $\tilde v$ is
{\em asymptotically regular of order s}
provided 
each of these $2^m$ tensor fields, 
multiplied by $\Omega^{-s}$, 
is smoothly extendible to ${\cal H}$.
Asymptotic regularity of a general tensor
field is defined 
by lowering any contra-variant indices with
$\tilde g_{ab}$ and applying the definition above
to the resulting covariant field.
Note that conditions $(1)(ii),(iii)$ above
are precisely the statement that $\tilde g_{ab}$ is asymptotically
regular of order 0; and
conditions $(3)(i)$--$(iii)$
are precisely the statement that $\tilde G_{ab}$ is asymptotically
regular of order 4.
The outer product of two asymptotically regular fields, 
of respective orders $s$ and $s'$,
is asymptotically regular, of order $s+s'$.
Contractions using $\tilde g^{ab}$
preserve asymptotic regularity, and order.

Thus, an asymptotically regular physical field 
gives rise, on $M$, to $2^m$ smooth fields, 
with ranks ranging from $m$ down to zero,
whose behavior near ${\cal H}$ reflects
the asymptotic behavior of the physical field.
Let $u_{a_1\cdots a_m}$ denote any one of these fields.
Then set, for $k$ any non-negative integer,
\begin{equation}
\met{u}{k}_{a_1\cdots a_m}
\equiv
\underleftarrow{\psi} 
\left [
({\cal L}_{(n\cdot\nabla\Omega)^{-1} n})^k 
u_{a_1\cdots a_m}
\right ]
, 
\label{rem}
\end{equation}
where $\underleftarrow{\psi}$ stands for the pull-back 
to ${\cal H}$ via the natural embedding map 
${\cal H}\stackrel{\psi}{\rightarrow} M$.
Note the right side of eqn.(\ref{rem}) exists since
$u_{a_1\cdots a_m}$ 
(and therefore each of its derivatives) is smoothly
extendible to ${\cal H}$.
The $\met{u}{k}{_{a_1\cdots a_m}}$ so defined will be 
called the {\em $k$-th order  remnant}\/ of $u{_{a_1\cdots a_m}}$.
These remnants, $(k=0,1,...)$, clearly carry, order by order, the asymptotic
information contained in $u_{a_1\cdots a_m}$,
and, therefore, the asymptotic information in the original physical 
field $\tilde v$.
Suppose, next that the physical field $\tilde v$ 
satisfies various field equations.
Then these field equations yield
partial differential equations 
on $M$ on the $u$'s that arise 
via asymptotic regularity from $\tilde v$,
and so partial differential equations
on ${\cal H}$ on the remnants $\met{u}{k}$ that
arise via eqn.(\ref{rem}) from the $u$'s.
We will refer to these as the {\em remnant field equations}.

We give some examples of asymptotically regular fields and
their associated remnants. 
Fix a space-time
$(\tilde M,\tilde g_{ab})$ with a completion $(M,\Omega )$.
For the first example,
consider the space-time metric $\tilde g_{ab}$.
Then, as we mentioned above, this field is asymptotically
regular of order 0. The corresponding $u$'s are
$ q_{ab} (\equiv \Omega^2 q_a{^k}q_b{^l}\tilde g_{kl})$,
$ 0 (\equiv \Omega^3 q_a{^k}n{^l}\tilde g_{kl})$,
and 
$ \lambda^{-2} (\equiv 
[\Omega^4 n^an^b \tilde g_{ab} ] )$.
Their corresponding remnants, 
$\met{q}{k}{_{ab}}$
and
$\met{\lambda}{k}$, carry the asymptotic
information contained in the space-time geometry.
We note that conditions
$(1)(iv)$ and $(1)(v)$ 
in the definition of a completion are
actually conditions on these remnants: namely
$\met{\lambda}{0}=1$, 
and $\met{q}{1}{_{ab}}=-2\met{\lambda}{1}\met{q}{0}{_{ab}}$
respectively.
For the second example, consider the Einstein tensor $\tilde G_{ab}$.
Then, as we mentioned above,
this field is asymptotically regular of order $4$. 
The corresponding $u$'s, written in terms of the
stress-energy tensor $\tilde T_{ab}$
( $= \tilde G_{ab} / \kappa$, with $\kappa = 8\pi G/c^4$)
are
\begin{equation}
T\equiv  \lambda^2 n^an^b (\kappa \tilde T_{ab}),\ \ \ 
T_a\equiv \lambda  \Omega^{-1} q_a{^k}n^l (\kappa \tilde T_{kl}),\ \ \ 
\mbox{\rm and}\ \ \ T_{ab}\equiv 
\Omega^{-2} q_a{^k}q_b{^l} 
\kappa ( \tilde T_{kl} -\frac{1}{2}\tilde T\tilde g_{kl}),
\label{teq}
\end{equation}
where we have 
introduced certain powers of $\lambda$ 
and have used the trace-reversed version of $\tilde T_{ab}$
in defining $T_{ab}$ for later convenience.
We denote by 
$\met{T}{k}$, 
$\met{T}{k}{_a}$, 
$\met{T}{k}{_{ab}}$, 
the remnants of $T$, $T_a$, and $T_{ab}$ respectively.
For the third example, 
consider the Weyl tensor, $\tilde C_{abcd}$,
of this space-time.
It is shown in Appendix C (c.f. the discussion
around eqn.(\ref{epot})) that this field is 
asymptotically regular of order $3$.
The $u$'s in this case may be taken to be
\begin{equation}
E_{ab}\equiv \Omega^{3} \lambda^2 q_a{^j}q_b{^l} n^k n^m\tilde C_{jklm},
\ \ \ \ 
\mbox{\rm and}
\ \ \ \ 
B_{ab}\equiv \Omega^{3} \lambda^2 q_a{^j}q_b{^l} n^k n^m{{^*}\tilde C}_{jklm}.
\label{weylebdef}
\end{equation}
Denote their remnants $\met{E}{k}{_{ab}}$ and $\met{B}{k}{_{ab}}$.
Note that 
condition $(1)(v)$ in the definition of a completion
is actually a condition on one of these remnants, namely
$\met{B}{0}{_{ab}}=0$
(c.f. eqn.(\ref{gr0.2b})). 
For the final example, consider a Maxwell field 
$\tilde F_{ab}$ in this space-time.
We demand that it be 
asymptotically regular of order $2$,$^{16}$
i.e., that each of 
\begin{equation}
E_a\equiv \Omega \lambda n^b\tilde F_{ab},\ \ \ 
\mbox{and}\ \ \ 
B_a\equiv \Omega \lambda n^b\,{{^*}\!\tilde F}_{ab}
\label{ebdef}
\end{equation}
be smoothly extendible to ${\cal H}$.
These are effectively the $u$'s.
This demand reflects the idea that 
a physically reasonable Maxwell field 
must fall off like $1/r^2$ near spatial infinity.
We denote by $\met{E}{k}{_a}$, $\met{B}{k}{_a}$ 
the remnants of $E_a$, $B_a$ respectively.
Note that it follows that the stress-energy tensor 
of this Maxwell field
has the fall-off rate 
consistent with that of eqn.(\ref{teq}).
Indeed, from
$\tilde T_{ab}=\frac{1}{2}(\tilde F_{am}\tilde F_b{^m}-\frac{1}{4}
\tilde F^2\tilde g_{ab} )$
we have 
\begin{equation}
T=\frac{1}{2}\kappa (E{^2} +B{^2}), \ \ \ 
{T}{_a}=-\kappa \epsilon_{amn}{E}{^m} {B}{^n}, \ \ \ 
{T}{_{ab}}= \kappa  [{E}{_a}{E}{_b}+{B}{_a}{B}{_b}
-\frac{1}{2}({E}{^2} +{B}{^2}){q}{_{ab}}]. 
\label{stress}
\end{equation}

There remains, as it turns out, some gauge freedom in the present framework. 
Fix a space-time $(\tilde M,\tilde g_{ab})$, 
and let $(M,\Omega )$ and  $(M,\Omega ')$, be two completions of 
$(\tilde M,\tilde g_{ab})$ 
It then follows that 
$\Omega '=\Omega (1+\omega\Omega )$, 
for some smooth function $\omega$ on $M$ such that
$\met{\omega}{0}\equiv \omega |_{\cal H}$ satisfies 
eqn.(\ref{tr0}) below
(i.e., $\met{\omega}{0}$ is an ``asymptotic translation'');
and, conversely, for $(M,\Omega )$ any completion and
$\omega$ 
and $\Omega '$ as above, then $(M,\Omega ')$ is also a completion. 
Thus the gauge freedom consists 
precisely of such $\omega$-fields.
The asymptotic gauge freedom, then, is described by the
remnants, $\met{\omega}{k}$, of $\omega$.
It turns out$^{8}$,
that one can, utilizing this gauge freedom, always achieve
\begin{equation}
\met{\lambda}{k} = 0, \ \ \ k\ge 2.
\label{gauge-fixing}
\end{equation}
and that this exhausts the gauge freedom associated with the remnants
$\met{\omega}{k}$, for $k\ge 1$.
Thus, making this gauge choice,
the remaining gauge freedom
is represented by a single $\met{\omega}{0}$
satisfying eqn.(\ref{tr0}).

\subsection{Asymptotic Translations}

In order to construct conserved quantities, it will be convenient
to have at hand some facts about asymptotic translations. 
Denote by ${\cal T}$ the set of functions $v$
on ${\cal H}$ satisfying the differential equation
\begin{equation}
D_aD_bv+v \met{q}{0}{_{ab}}=0,
\label{tr0}
\end{equation}
where $D_a$ denotes the derivative operator of $\met{q}{0}{_{ab}}$.
This ${\cal T}$ is a 4-dimensional vector space 
(since, by virtue of the fact that the curl of 
eqn.(\ref{tr0}) is an identity, $v$ is completely determined 
by its value and derivative at any one point)
equipped with a Lorentz metric
$\langle v,w\rangle \equiv
\met{q}{0}{^{ab}}D_av D_bw +vw $ 
(since, by virtue of eqn.(\ref{tr0}), 
the right side is a constant ).
Elements of ${\cal T}$ can be 
interpreted$^{11}$ as asymptotic translations on $M$
in the following sense: For $\tilde\xi^a$ a vector field
on $\tilde M$
asymptotically regular of order $0$ such that 
${\cal L}_{\tilde\xi}\tilde g_{ab}$ is asymptotically regular
of order $2$,
then $\Omega^{-2}{\cal L}_{\tilde \xi}\Omega |_{\cal H}\in {\cal T}$.

It is convenient to introduce an index notation for tensors
over ${\cal T}$: Greek superscripts and subscripts 
denote, respectively, elements of ${\cal T}$ and 
its dual ${\cal T}^*$.
Thus a solution $v$ of eqn.(\ref{tr0}) might be
denoted $v^\mu$, while 
a linear map ${\cal T}\stackrel{w}{\mapsto}\mathbb{R}$
might be denoted $w_{\mu}$.
The action of $w$ on $v$ would be expressed by contraction:
$w(v)=w_\mu v^\mu$.
We denote by $\eta_{\mu\nu}$ the above Lorentz metric on ${\cal T}$,
i.e., we set $\eta_{\mu\nu}v^\mu w^\nu =\langle v,w\rangle$.
We shall use $\eta_{\mu\nu}$ (and its inverse) 
to lower and raise indices of tensors over ${\cal T}$.
The objects with which we shall be concerned are fields on ${\cal H}$
that may have Latin indices (indicating tensor character over the
manifold ${\cal H}$) and Greek indices (indicating tensor character over 
the vector space ${\cal T}$). Thus, for example, $\zeta_\alpha$ would
denote a ${\cal T}^*$-valued function on ${\cal H}$,
$\zeta^a$ would denote an ordinary tangent vector field on ${\cal H}$,
and $\zeta^a{_\alpha}$ would denote a ${\cal T}^*$-valued
tangent vector field on ${\cal H}$. 
In particular, an element $v^\mu$ in ${\cal T}$
is now viewed as a ${\cal T}$-valued constant function on ${\cal H}$.
We lower and raise
Greek indices of such fields with $\eta_{\alpha\beta}$ and its
inverse, and lower and raise Latin indices with $\met{q}{0}{_{ab}}$
and its inverse.
There is a natural field, $\alpha_\mu$, defined by
the property that,
for any $v^\mu\in {\cal T}$, $\alpha_\mu v^\mu$ is the corresponding
solution of eqn.(\ref{tr0}).
Then, e.g., $\alpha_\mu\alpha^\mu = 1$. The derivative operator
$D_a$ on ${\cal H}$ associated with $\met{q}{0}{_{ab}}$ extends to
a derivative operator on our indexed fields by demanding that
$D_av^\alpha = 0$, for
$v^\alpha $ any constant field. 
There now follows
$D_a \met{q}{0}{_{bc}} = 0$,
$D_a \eta_{\alpha\beta} = 0$,
\begin{equation}
D_aD_b\alpha_{\mu}+\alpha_\mu\met{q}{0}{_{ab}}=0
\label{tr}
\end{equation}
(from eqn.(\ref{tr0})), 
\begin{equation}
D_a\alpha_\mu D^a\alpha_\nu +\alpha_\mu\alpha_\nu 
=\eta_{\mu\nu}
\label{tr1}
\end{equation}
(from the definition of $\eta_{\mu\nu}$),
and $\eta^{\mu\nu} D_a\alpha_\mu D_b\alpha_\nu =\met{q}{0}{_{ab}}$.$^{17}$

\subsection{Conserved Quantities}

Now imagine that we were somehow able to find a 
divergence-free vector field, 
constructed from (the remnants of) some physical
fields and the background geometry of ${\cal H}$.
Integrating (the dual of) this 
vector field over
a 2-sphere cut (i.e., a non-contractible 2-sphere
sub-manifold) of ${\cal H}$, we obtain
a number---one clearly independent of choice of cut.
Think of such an integral as being the limit of an integral over 
a space-like 2-sphere in space-time, as the 2-sphere
approaches the cut at spatial infinity. 
These integrals we call conserved quantities.
In each of the examples we shall consider, 
the divergence-free vector field
is multi-linear in $\alpha_\mu$,$^{18}$
and so the conserved quantities may be viewed as a tensor over ${\cal T}$.

We now give three well-known$^{1, 2, 4, 11}$
examples of conserved quantities.
Some of the computations are relegated to section IV and Appendix C.
Fix a space-time $(\tilde M,\tilde g_{ab})$, 
a completion $(M,\Omega )$ thereof and a cut $C$ of ${\cal H}$.

For the first  example, 
let $\tilde F_{ab}$ be a Maxwell field on $\tilde M$,
regular of order 2.
Consider the right side of 
\begin{equation}
Q=\frac{1}{4\pi}\int_C 
\met{E}{0}{^a} dS_a ,
\label{charge}
\end{equation}
where $\met{E}{0}{^a}$ is the (zero-th order) remnant of
$E_a$ given by eqn.(\ref{ebdef}).
Maxwell's equations imply the integrand above is divergence-free
(c.f. eqn.(\ref{em0.1})). 
Thus eqn.(\ref{charge}) defines a
conserved quantity. 
This $Q$ is precisely the electric charge, 
for the right side of
eqn.(\ref{charge}) is the limit of the integral 
of ${^*F_{ab}}$ over a large space-like 2-sphere in the space-time
as that 2-sphere approaches the cut $C$. 
For the second  example, 
consider the right side of
\begin{equation}
{\cal P}^{\mu}
=\frac{1}{8\pi}
\int_C
\met{E}{0}{^{ab}}
D_b\alpha^\mu dS_a,
\label{4m}
\end{equation}
where $\met{E}{0}{_{ab}}$
is the remnant of $E_{ab}$,
a portion of the Weyl tensor,
 given in eqn.(\ref{weylebdef}).
The remnant field equation (eqn.(\ref{gr0.1}))
together with eqn.(\ref{tr}) on $\alpha_\mu$, 
imply that the integrand above is divergence-free 
Thus eqn.(\ref{4m}) defines a conserved quantity,
which is a vector over ${\cal T}$.
This ${\cal P}^\mu$ 
is precisely$^{1,2}$  
the total mass-momentum of the space-time.
For the third example, 
consider the right side of
\begin{equation}
{\cal M}_{\mu\nu}
=-\frac{1}{16\pi}
\epsilon_{ \mu\nu } {^{\tau\sigma }}
\int_C
\met{B}{1}{^{ab}}
\alpha_{\tau}
D_b\alpha_{\sigma } dS_a,
\label{angdef}
\end{equation}
where $\epsilon_{ \mu\nu \tau\sigma }$
denotes the $\eta$-alternating tensor on ${\cal T}$. 
In order for the integrand above to be divergence-free,
we must impose on the space-time the additional condition$^{19}$
that 
\begin{equation}
D_{[a}\met{T}{0}{_{b]}}=0.
\label{angcd}
\end{equation}
Under this additional condition,
 eqn.(\ref{angdef}) defines a conserved quantity,
which is a two-form over ${\cal T}$.
This ${\cal M}_{\mu\nu}$ 
is precisely$^{4}$
the total angular momentum of the space-time.

Finally, we revisit the issue of gauge.
Fix a space-time $(\tilde M,\tilde g_{ab})$,
and a completion $(M,\Omega )$ thereof.
Demand further that the completion
satisfy the gauge condition (\ref{gauge-fixing}),
so the remaining gauge freedom is represented by
the choice of some
$\met{\omega}{0}\in {\cal T}$.
Applying such a gauge transformation, 
the remnants of any physical field, and thus also 
of any 
conserved quantities
associated with that field, will in general change.
Specifically, let $Q_{A}$
be any conserved quantity 
or any remnant field, where the subscript $A$ is an
abbreviation for all the indices of $Q$.
Then, for each translation $\met{\omega}{0}\in {\cal T}$,
there corresponds a 
``gauge-transformed'' 
quantity---$Q_{A}[\met{\omega}{0}]$.
Thus, we may regard our quantity
$Q_{A}$
as a tensor field on the 4-manifold ${\cal T}$ so
defined that its value at
$\met{\omega}{0}\in {\cal T}$ is $Q_{A}[\met{\omega}{0}]$.
In short, 
the gauge behavior of 
our original quantity
$Q_{A}$
is coded in the position dependence of this tensor field
on ${\cal T}$.
The derivative of this tensor field reflects the
behavior of the quantity under 
``infinitesimal gauge transformation''.
Indeed, from
\begin{equation}
Q_{A}[
\met{\omega}{0}
+ \delta \met{\omega}{0}
]
= Q_{A}[ \met{\omega}{0} ] +
(\delta\met{\omega}{0}){^\mu}
Q^{(1)} {_{\mu A}} [ \met{\omega}{0} ] + 
O((\delta\met{\omega}{0})^2),
\end{equation}
we have
\begin{equation}
\nabla_\mu Q_{A}
= Q^{(1)} {_{\mu A}},
\end{equation}
where $\nabla_\mu$ denote the natural 
derivative operator on the 4-manifold ${\cal T}$.
As examples, consider the conserved quantities
(\ref{charge})--(\ref{angdef}).
Under a gauge transformation,
$\Omega '=\Omega (1+\omega \Omega )$
with $\met{\omega}{0}\in {\cal T}$
the remnants $\met{E}{0}{_a}$,
$\met{E}{0}{_{ab}}$ remain
unchanged, while
$\met{B}{1}{_{ab}}$ 
changes to
$\met{B'}{1}{_{ab}}=
\met{B}{1}{_{ab}}
-2 \epsilon_{(a}{^{kl}}\met{E}{0}{_{b)k}}D_l\met{\omega}{0} $.
In terms of the corresponding tensor fields on the
4-manifold ${\cal T}$, these become,
$\nabla_\mu\met{E}{0}{_a} = 0$,
$\nabla_\mu\met{E}{0}{_{ab}} = 0$, and
$\nabla_\mu\met{B}{1}{_{ab}}=
-2 \epsilon_{(a}{^{kl}}\met{E}{0}{_{b)k}}D_l\alpha_\mu $.
It follows that the total electric charge $Q$ (\ref{charge})
and the 4-momentum ${\cal P}^\mu$ (\ref{4m})
are gauge invariant, and that$^{11}$
the angular momentum 
${\cal M}_{\mu\nu}$ (\ref{angdef})
changes via$^{20}$
\begin{equation}
{\cal M}'{_{\mu\nu}} =
{\cal M}_{\mu\nu} 
-\met{\omega}{0}{_{[\mu}} {\cal P}_{\nu ]}.
\end{equation}
In terms of the corresponding tensor fields on the
4-manifold ${\cal T}$, these become,
respectively,
$\nabla_\lambda Q = 0$,
$\nabla_\lambda {\cal P}^\mu = 0$,
and
\begin{equation}
\nabla_{\lambda } {\cal M}_{\mu\nu} =
-\eta_{\lambda  [\mu}{\cal P}_{\nu ]}.
\label{anggg}
\end{equation}

\section{MINKOWSKI SPACE-TIME}
We now apply the framework developed in the previous section
to the study of conserved quantities associated with physical fields
in  Minkowski space-time.
Minkowski space-time is a good starting point: 
It is simple, and suggestive
of what might happen in the presence of curvature.
We shall take as the physical field successively a Klein-Gordon field,
a Maxwell field and a linearized gravitational field.
We will write down, for each of these cases,
 all conserved quantities linear in 
the physical fields and multi-linear in asymptotic translations.

Let $(\tilde M, \tilde\eta_{ab})$ be Minkowski space-time.
Fix a point $p\in \tilde M$, 
let $\Omega$ be the inverse geodesic distance from $p$. 
Then this $\Omega$ yields a 
completion $(M,\Omega)$ of Minkowski space-time
which we call the {\em standard completion}. 
In this  completion, we have
$\met{\lambda}{n}=0$
and $\met{q}{n}{_{ab}}=0$, for $n\ge 1$.

\subsection{Remnant field equations}
Here we derive the remnant field equations for Klein-Gordon, 
Maxwell, and linearized gravitational fields for later use in constructing
conserved quantities.
For what follows we fix
a standard completion of Minkowski space-time $\tilde M$
and denote by $D_a$ the derivative operator associated with
$q_{ab}$ on constant-$\Omega$ surfaces.

Let $\tilde \phi$ be a Klein-Gordon field on $\tilde M$
asymptotically regular of order $1$.
Setting $\phi =\Omega^{-1}\tilde\phi$, 
we have
\begin{equation}
0=\tilde \nabla^2\tilde\phi=\Omega^3 
\left [
(D^2-1)\phi 
+\Omega {\cal L}_n \phi
+\Omega^2 ({\cal L}_n )^2\phi 
\right ].
\label{phimin}
\end{equation}
Taking the remnants of the above equation, we obtain,
\begin{equation}
D^2 \metbar{\phi}{n}= (- n^2 + 1) \metbar{\phi}{n},
\label{phin}
\end{equation}
for $n = 0,1,2,...$\ .

Let $\tilde F_{ab}$ be a Maxwell field on $\tilde M$ asymptotically regular
of order $2$, with remnants $\met{E}{n}{_a}$ and $\met{B}{n}{_a}$.
Using eqn.(\ref{ebdef}), Maxwell's equation can be written as
\begin{eqnarray}
0&=&\tilde \nabla^m \tilde F_{ma}=\Omega D^m E_m\nabla_a\Omega
-\Omega^2 (\Omega {\cal L}_n E_a -\epsilon_{akl}D^k B^l ),
\\
0&=&\tilde \nabla^m {^{*}\tilde F}_{ma}
=\Omega D^m B_m\nabla_a\Omega 
-\Omega^2 (\Omega {\cal L}_n B_a +\epsilon_{akl}D^k E^l ).
\label{maxmin}
\end{eqnarray}
Taking the remnants of the above equations, we obtain
\begin{xalignat}{2}
D_a \metbar{E}{0}{^a}&=0, & D_a 
\metbar{B}{0}{^a}&=0, 
\label{max0} \\
\epsilon^{abc}D_b \metbar{B}{n}_c &= n\metbar{E}{n}{^a}, &
-\epsilon^{abc}D_b \metbar{E}{n}_c &= n\metbar{B}{n}{^a},
\label{maxn}
\end{xalignat}
for $n = 0,1,2,...$\ .
Note that eqn.(\ref{maxn}) imply
\begin{xalignat}{2}
D^2 \met{E}{n}{_a}&=( - n^2 + 2 )\met{E}{n}{_a}, &
D^2 \met{B}{n}{_a}&=( -n^2 +2)\met{B}{n}{_a},
\label{d2M}
\end{xalignat}
for $n = 0,1,2,...$\ .

Let $\tilde K_{abcd}$ be a
linearized gravitational field on $\tilde M$, i.e.,
a tensor field on $\tilde M$ having the same symmetry 
and contractions as
the Weyl tensor and satisfying the 
linearized Bianchi identity:
\begin{equation}
\tilde\nabla_{[a}\tilde K_{bc]de}= 0 .
\end{equation}
Let $\tilde K_{abcd}$ be asymptotically regular of order $3$, 
so $E_{ab}\equiv\Omega^3\tilde K_{akbl}n^k n^l$ and
$B_{ab}\equiv\Omega^3{{^*}\tilde K}_{akbl}n^k n^l$ 
are smoothly extendible to ${\cal H}$. 
Their remnants, denoted $\met{E}{n}{_{ab}}$ and $\met{B}{n}{_{ab}}$,
are symmetric and  trace-free. The linearized Bianchi
identity can be written as
\begin{eqnarray}
0&=&\tilde\nabla^m\tilde K^{*}_{mabc} \nonumber \\
&=& 
\left [
\Omega^{-1}\nabla_a\Omega D^m B_{m[b}
-(\Omega {\cal L}_n B_{a[b}+\epsilon_{akl}D^k E^l{_{[b}} )
\right ]
\nabla_{c]}\Omega \nonumber \\
&&\ \ \ +\frac{1}{2}
\left [
\nabla_a\Omega D^k E_{km}
-\Omega (\Omega {\cal L}_n E_{ma}-\epsilon_{mkl}D^k B^l{_{a}} )
\right ]
\epsilon^m{_{bc}} .
\end{eqnarray}
Taking the remnants of the above equation, we obtain
\begin{xalignat}{2}
\epsilon^{lma}D_l \metbar{B}{n}_m{^{b}} &= n\metbar{E}{n}{^{ab}}, &
-\epsilon^{lma}D_l \metbar{E}{n}_m{^{b}} &= n\metbar{B}{n}{^{ab}},
\label{grn}
\end{xalignat}
for $n = 0,1,2,...$\ .
Note that, eqn.(\ref{grn}) imply
\begin{xalignat}{2}
D^2 \met{E}{n}{_{ab}}&=(-n^2 +3)\met{E}{n}{_{ab}}, &
D^2 \met{B}{n}{_{ab}}&=(-n^2 +3)\met{B}{n}{_{ab}},
\label{d2G}
\end{xalignat}
for $n = 0,1,2,...$\ .

Recall that the present framework is subject to a class
of restricted gauge transformations
(namely, replacements of $\Omega$ by
$\Omega '=\Omega (1+\omega\Omega )$),
which preserve the gauge conditions
$\met{\lambda}{n}=0$, $n\ge 2$,
and that each such gauge transformation is completely
characterized by an
$\met{\omega}{0}\in {\cal T}$.
For completeness, we summarize the behavior of the remnants
above under such a gauge transformation:
\begin{eqnarray}
\nabla_\mu 
\metbar{\phi }{n}&=&
n
(
{\cal L}_{D\alpha_\mu} 
\metbar{\phi}{n-1}
-n\alpha_\mu\metbar{\phi}{n-1}
)
\label{eq2}
\\
\nabla_\mu 
\met{E}{n}{_a}&=&
n
(
{\cal L}_{D\alpha_\mu}
\met{E}{n-1}{_a}
-n\alpha_\mu\metbar{E}{n-1}{_a}
+\epsilon_{a}{^{kl}}\met{B}{n-1}{_k}D_l\alpha_\mu
),
\label{eg} \\
\nabla_\mu 
\met{B}{n}{_a}&=&
n
(
{\cal L}_{D\alpha_\mu} 
\met{B}{n-1}{_a}
-n\alpha_\mu \metbar{B}{n-1}{_a}
-\epsilon_{a}{^{kl}}\met{E}{n-1}{_k}D_l\alpha_\mu
),
\\
\nabla_\mu
\met{E}{n}{_{ab}}&=&
n
(
{\cal L}_{D\alpha_\mu} \met{E}{n-1}{_{ab}}
-n\alpha_\mu\metbar{E}{n-1}{_{ab}}
+2\epsilon^{kl}{_{(a}}\met{B}{n-1}{_{b)k}}D_l\alpha_\mu
),
\label{geg} \\
\nabla_\mu
\met{B}{n}{_{ab}}&=&
n
(
{\cal L}_{D\alpha_\mu}\met{B}{n-1}{_{ab}}
-n\alpha_\mu \met{B}{n-1}{_{ab}}
-2\epsilon^{kl}{_{(a}}\met{E}{n-1}{_{b)k}}D_l\alpha_\mu
),
\end{eqnarray}
for $n = 0,1,2,...$\ .
Note that $\metbar{\phi}{0}$,
$\met{E}{0}{_a}$, 
$\met{B}{0}{_a}$,
$\met{E}{0}{_{ab}}$ 
and $\met{B}{0}{_{ab}}$ 
are gauge invariant. 
As a consistency check, we note 
also that the $\nabla_\mu$-curl of the right side
of each of the above equation vanishes, by virtue of 
$\nabla_\mu \alpha_\nu = 0$, as it must.
Of course, these gauge-transformed fields satisfy the same equations
as the original fields.

\subsection{Remnant Radiation Multipoles}
It is perhaps most natural to seek conserved quantities
that are linear in the remnants,
since, e.g.,
this category includes all well-known conserved quantities.$^{21}$
In this section we shall find all such conserved quantities
for a Klein-Gordon field, a Maxwell field and
a linearized gravitational field in Minkowski space-time $\tilde M$. 
Again, we fix the standard completion of $\tilde M$.

We begin with the Klein-Gordon field.
Let $\tilde \phi$ be a Klein-Gordon field asymptotically
regular of order $1$, with remnants $\met{\phi}{n}$.
\begin{theorem}
$(i)$ The conserved quantities linear in this Klein-Gordon field
consist precisely of
the family
\begin{equation}
{\cal K}_{\mu_1\cdots \mu_{n-1}}
[ \met{\phi}{n} ]
\equiv
\int_C
\left [
{\cal C}(\alpha_{\mu_1} \cdots\alpha_{\mu_{n-1}})
D^a\metbar{\phi}{n}
- \metbar{\phi}{n}
D^a 
{\cal C}(\alpha_{\mu_1} \cdots\alpha_{\mu_{n-1}})
\right ] dS_a 
,\ n\ge 1, \label{K}
\end{equation}
where
${\cal C} (\alpha_{\mu_1} \cdots \alpha_{\mu_{n-1}})$
denotes the symmetric, trace-free part of
$\alpha_{\mu_1} \cdots \alpha_{\mu_{n-1}}$.$^{22}$

\noindent $(ii)$ The ${\cal K}_{\mu_1\cdots\mu_{n-1}}$ are totally
symmetric and trace-free.

\noindent $(iii)$ The behavior of ${\cal K}_{\mu_1\cdots\mu_{n-1}}$
under restricted gauge transformations is given by
\begin{equation}
\nabla_{\mu} {\cal K}_{\mu_1\cdots\mu_{n-1} }
=
\frac{1}{2}n(n-2)\eta_{(\mu_1\mu_2}
{\cal K}_{\mu_3\cdots\mu_{n-1} )\mu}-
n(n-1)\eta_{\mu (\mu_1}
{\cal K}_{\mu_2\cdots\mu_{n-1} )}.
\label{mult}
\end{equation}
\end{theorem}
\noindent We will refer to these ${\cal K}$'s
as the remnant radiation multipoles of a Klein-Gordon field.

To see that eqn.(\ref{K}) indeed defines a conserved quantity, 
take the divergence of the integrand, and use that both 
$\met{\phi}{n}$ and 
    ${\cal C} (\alpha_{\mu_1} \cdots \alpha_{\mu_{n-1}})$
    satisfy eqn.(\ref{phin}). 
To prove $(iii)$, use
eqn.(\ref{eq2}), the
    definition of ${\cal K}$, and a certain identity on
    $ {\cal C}(\alpha_{\mu_1} \cdots\alpha_{\mu_{n-1}})$.
    See Appendix B for details.
 Note the right side of eqn.(\ref{mult}) is, up to 
an overall factor,
    the only 
$(n-1)$-th rank, symmetric, trace-free tensor linear in
$\met{\cal{K}}{n-2}$.
    Eqn.(\ref{mult}) states that the 
    dependence of the ${\cal K}$'s on position in ${\cal T}$
    is exactly that of ordinary multipole moments.
The proof that the family
given by eqn.(\ref{K})
exhausts the linear
conserved quantities in the Klein-Gordon case is outlined in Appendix B.
 As an example of these remnant radiation multipoles,
    let $\tilde\phi =(f(t+r)-f(t-r))/r$. Then, provided
    $k_\pm (x)\equiv f(\pm \frac{1}{x}), x>0$, are
    both smoothly
    extendible to zero, this $\tilde\phi$
    will be asymptotically regular of order 1.
    Then the remnants of $\tilde\phi$ are given by
    $\met{\phi}{n}=(1+\zeta^2)^{-1/2}\{
    k^{(n)}_+(0)[(1+\zeta^2)^{1/2}-\zeta ]^n -
    k^{(n)}_-(0)[(1+\zeta^2)^{1/2}+\zeta ]^n \}$,
where we have set $\zeta = -\Omega^{-2}\frac{\partial \Omega }
    {\partial t} |_{\cal H} \in {\cal T}$.
    The ${\cal K}$'s in this example involve
    various derivatives of $k_{\pm}$ at zero.
    Explicitly, the first two are given by
 ${\cal K}=4\pi [k'_+(0) +k'_-(0)]$,
    ${\cal K}_\mu =4\pi [k^{''}_-(0) +k^{''}_-(0)]
    \langle \alpha_\mu ,\zeta \rangle $.
    Thus the ${\cal K}$'s in this example describe 
    radiation emanating from future and past time-like infinity.

We turn next to the Maxwell case.
Let $\tilde F_{ab}$ be a Maxwell field asymptotically regular of
order 2, with remnants $\met{E}{n}{_a}$ and $\met{B}{n}{_a}$.
\begin{theorem}
$(i)$ The conserved quantities linear in this Maxwell field
consist precisely of the electric charge 
(given by eqn.(\ref{charge})),
the magnetic charge (obtained by replacing 
$\met{E}{0}{_a}$
by
$\met{B}{0}{_a}$
in eqn.(\ref{charge})), and
the family
\begin{eqnarray}
{\cal E}_{\mu\mu_1\cdots\mu_{n-1}}
&\equiv &
{\cal K}_{\mu_1\cdots\mu_{n-1}} [\met{E}{n}{^m}D_m\alpha_\mu ]
\nonumber \\
&=&\int 
\left [
D^a (\met{E}{n}{^m}D_m\alpha_\mu )
{\cal C}(\alpha_{\mu_1}
\cdots \alpha_{\mu_{n-1}} )
-\ \met{E}{n}{^m}D_m\alpha_\mu 
D^a  {\cal C}(\alpha_{\mu_1}
\cdots \alpha_{\mu_{n-1}} )
\right ]
dS_a ,
\label{caledef}
\end{eqnarray}
for $n= 1,2,3,...$\ .

\noindent $(ii)$ 
The 
${\cal E}_{\mu\mu_1\cdots\mu_{n-1}}$ are trace-free
in all indices,
totally symmetric
in the indices $\mu_1\cdots\mu_{n-1}$,
and satisfy
\begin{equation}
{\cal E}_{(\mu\mu_1\cdots\mu_{n-1})}=0.
\label{esym}
\end{equation}

\noindent $(iii)$ The 
gauge behavior of ${\cal E}_{\mu\mu_1\cdots\mu_{n-1}}$ is given by
\begin{eqnarray}
\nabla^{\mu} {\cal E}^{\nu}{_{\mu_1\cdots\mu_{n-1} }}
&=&
\frac{1}{2}n(n-2)\eta_{(\mu_1\mu_2}
{\cal E}^{\nu\mu }{_{\mu_3\cdots\mu_{n-1} )}}-
n(n-1)\delta^{\mu }{_{ (\mu_1}}
{\cal E}^{\nu }{_{ \mu_2\cdots\mu_{n-1} )}}
\nonumber \\
&&\ +
\frac{n(n-2)}{n-1}\eta_{(\mu_1\mu_2}
{\cal E}^{[\nu\mu ]}{_{\mu_3\cdots\mu_{n-1} )}}
- 2n\delta^{[\mu }{_{ (\mu_1}}
{\cal E}^{\nu ]}{_{ \mu_2\cdots\mu_{n-1} )}}.
\label{multe}
\end{eqnarray}

\label{Max}
\end{theorem}

\noindent We will refer to the ${\cal E}$'s
as the remnant radiation multipoles of a Maxwell field.

To see that eqn.(\ref{caledef}) indeed defines
a conserved quantity, take the divergence of the integrand
and use that
$\met{E}{n}{^m}D_m\alpha_\mu $ 
(and $\met{B}{n}{^m}D_m\alpha_\mu$)
satisfy eqn.(\ref{phin}).
To prove eqn.(\ref{esym}), we note that its integrand
is the divergence of an anti-symmetric tensor.$^{23}$
Eqn.(\ref{esym}) implies, in particular, that
${\cal E}_\mu$ is zero, 
and that
${\cal E}_{\mu\nu}$ is anti-symmetric. 
While a second family of conserved quantities,
${^*}{\cal E}$,
associated similarly with $\met{B}{n}{_a}$ could
be defined, they yield nothing new, for we have$^{24}$
\begin{equation}
{^{*}{\cal E}}_{\mu\mu_1\cdots\mu_{n-1}}
= \frac{n-1}{n}\epsilon_{\mu (\mu_1}{^{\nu\nu_1}}
{\cal E}_{|\nu\nu_1 |\mu_2\cdots\mu_{n-1})} .
\label{edual}
\end{equation}
Note that
${^{*}{\cal E}}_{\mu\mu_1\cdots\mu_{n-1}}$ 
has the symmetries $(ii)$ in theorem \ref{Max} above and that 
${^{**}{\cal E}}_{\mu\mu_1\cdots\mu_{n-1}}
=-{\cal E}_{\mu\mu_1\cdots\mu_{n-1}}$.
The gauge behavior, eqn.(\ref{multe}), is proved in Appendix B.
Note eqn.(\ref{multe}) yields, in particular, that
${\cal E}_{\mu\nu }$ is gauge invariant.
The proof that the quantities given
by eqn.(\ref{caledef}) exhaust the 
linear, Maxwell conserved quantities is outlined in
Appendix B.
Here is an example of these electro-magnetic
conserved quantities. Let
$\tilde\phi$ be a Klein-Gordon field asymptotically regular
of order 1, $\tilde w^{ab}$ a constant anti-symmetric tensor
field on $\tilde M$, and set $\tilde F_{ab} = \tilde \nabla_{[a}
(\tilde w_{b]m}\tilde\nabla^m\tilde\phi )$.
Then this $\tilde F_{ab}$ is a solution of Maxwell's equations,
asymptotically regular of order 2. Its remnants are given
in terms of those of $\tilde\phi$ by
\begin{equation}
\met{E}{n}{_a}=D_aD_b\met{\phi}{n-1}\xi^b -(n+1)D_m\xi_a D^m\met{\phi}{n-1}
+n^2\met{\phi}{n-1}\xi_a,
\end{equation}
where we have set 
$\xi^a = \tilde w^{ab}x_b $.
Then, the remnant radiation multipoles of $\tilde F_{ab}$
can be expressed in terms of those of $\tilde\phi$.
For instance, we have
${\cal E}_{\mu\nu}={\cal K}w_{\mu\nu}$, where
we have set
$w_{\mu\nu}\equiv 2 \xi_a \alpha_{[\mu}D^a\alpha_{\nu ]}
+ D_a\xi_b D^a\alpha_\mu D^b\alpha_\nu$.

We turn finally to linearized gravity.
Let $\tilde K_{abcd}$ be a linearized gravitational field
asymptotically regular of
order 3, with remnants $\met{E}{n}{_{ab}}$ and $\met{B}{n}{_{ab}}$.
\begin{theorem}
$(i)$ 
The conserved quantities linear in this linearized gravitational field
consist precisely of
the mass-momentum (given by eqn.(\ref{4m})), 
the angular momentum, (given by eqn.(\ref{angdef})), 
and 
\begin{eqnarray}
{\cal G}_{\mu\nu\mu_1\cdots\mu_{n-1}}
&\equiv &
{\cal K}_{\mu_1\cdots\mu_{n-1}} [\met{E}{n}{^{kl}}
D_k\alpha_\mu 
D_l\alpha_\nu ]
= {\cal E}_{\nu\mu_1\cdots\mu_{n-1}}[\met{E}{n}{_{ab}}D^b\alpha_\mu ] 
\nonumber \\
&=&\int 
\left [
D^a 
(\met{E}{n}{^{kl}} D_k\alpha_\mu D_l\alpha_\mu )
{\cal C}(\alpha_{\mu_1}
\cdots \alpha_{\mu_{n-1}} )
\right .
\nonumber \\
&&\hspace{1.2cm}
\left .-\ \met{E}{n}{^{kl}} D_k\alpha_\mu D_l\alpha_\mu 
D^a  {\cal C}(\alpha_{\mu_1}
\cdots \alpha_{\mu_{n-1}} )
\right ] dS_a ,
\label{calgdef}
\end{eqnarray}
for $n=1,2,3,...$\ .

\noindent $(ii)$ 
The ${\cal G}_{\mu\nu\mu_1\cdots\mu_{n-1}}$
are trace-free in all indices, 
totally symmetric in the indices $\mu_1\cdots\mu_{n-1}$, 
symmetric in indices $\mu$, $\nu$,
and satisfy
\begin{equation}
{\cal G}_{\mu (\nu\mu_1\cdots\mu_{n-1})}=0.
\label{gsym}
\end{equation}

\noindent $(iii)$ The 
gauge behavior of 
${\cal G}_{\mu\nu\mu_1\cdots\mu_{n-1}}$
is given by
\begin{eqnarray}
\nabla^{\sigma} {\cal G}^{\mu\nu}{_{\mu_1\cdots\mu_{n-1} }}
&=&
\frac{1}{2}n(n-2)\eta_{(\mu_1\mu_2}
{\cal G}^{\mu\nu\sigma  }{_{\mu_3\cdots\mu_{n-1} )}}-
n(n-1)\delta^{\sigma }{_{ (\mu_1}}
{\cal G}^{\mu\nu }{_{ \mu_2\cdots\mu_{n-1} )}}
\nonumber \\
&&\ +
\frac{n(n-2)}{n-1}\eta_{(\mu_1\mu_2}
( {\cal G}^{\mu [\nu\sigma ]}{_{\mu_3\cdots\mu_{n-1} )}}
+ {\cal G}^{\nu [\mu\sigma ]}{_{\mu_3\cdots\mu_{n-1} )}} )
\nonumber \\
&&\ - 2n\delta^{[\sigma }{_{ (\mu_1}}
({\cal G}^{\mu ]\nu}{_{ \mu_2\cdots\mu_{n-1} )}}
+{\cal G}^{\nu ]\mu}{_{ \mu_2\cdots\mu_{n-1} )}} ) .
\label{multg}
\end{eqnarray}

\label{Grav}
\end{theorem}

\noindent We will refer to the ${\cal G}$'s
as the remnant radiation multipoles of a linearized gravitational field.

To see that eqn.(\ref{calgdef}) indeed defines 
a conserved quantity, 
take the divergence of the integrand and use that
$\met{E}{n}{^{kl}} D_k\alpha_\mu D_l\alpha_\nu $ 
(and $\met{B}{n}{^{kl}} D_k\alpha_\mu D_l\alpha_\nu $)
satisfy eqn.(\ref{phin}).
Eqn.(\ref{gsym}), which is actually equivalent to
${\cal E}_{(\nu\mu_1\cdots\mu_{n-1})}=0$,
implies, in particular, that
${\cal G}_{\mu\nu}={\cal G}_{\mu\nu\sigma}=0$. 
While a second family of conserved quantities, ${^*}{\cal G}$,
associated similarly with $\met{B}{n}{_{ab}}$ could
be defined, they yield nothing new, for we have$^{25}$
\begin{equation}
{^{*}{\cal G}}{^{\mu\nu}}{_{\mu_1\cdots\mu_{n-1}}}
= \frac{n-1}{n}
\epsilon^{\sigma (\mu }{_{\lambda (\mu_1}} 
{\cal G}^{\nu)\lambda}{_{|\sigma |\mu_2\cdots\mu_{n-1})}}.
\label{gdual}
\end{equation}
Note that
${^{*}{\cal G}}_{\mu\nu\mu_1\cdots\mu_{n-1}}$ 
also satisfies $(ii)$ in theorem \ref{Grav} and that
${^{**}{\cal G}}_{\mu\nu\mu_1\cdots\mu_{n-1}}
=-{\cal G}_{\mu\nu\mu_1\cdots\mu_{n-1}}$. 
The proof that
the quantities given by
eqn.(\ref{calgdef})
exhaust the linear, gravitational conserved quantities is
outlined in Appendix B.
We omit the proof of the gauge behavior (eqn.(\ref{multg})), 
which is similar to the Maxwell case.
Examples of linearized gravitational fields,
their remnants, and their remnant radiation
multipoles can be 
constructed in a manner similar to that of the Maxwell case.$^{26}$

One might expect, on physical grounds, 
that a static field
would be characterized completely by its static multipole moments
and that its remnant radiation multipoles 
would all vanish identically. 
This indeed turns out to be the case. See Appendix A.

\section{CURVED SPACE-TIME}
It is natural to ask whether 
the remnant radiation multipoles constructed above for various fields
in Minkowski space-time 
can be generalized to curved space-time. 
%
%
%
%
%
%
To address this issue,
we
first obtain the remnant equations.
Let $\tilde \phi$ be a Klein-Gordon field 
asymptotically regular of order $1$,
so,
$ \phi ( \equiv \Omega^{-1}\tilde\phi )$ is
smoothly extendible to ${\cal H}$.
Then the Klein-Gordon equation on $\tilde\phi$ yields 
\begin{eqnarray}
0=\tilde \nabla^2\tilde\phi &=&\Omega^3 [
D^2\phi 
+\lambda^{-1} D^a\lambda D_a\phi 
+ \Omega \lambda^{-2}(2\met{\phi}{1} + \Omega \met{\phi}{2} )
\nonumber \\ &&\ \ \ \ \ \ 
+ (\phi + \Omega \met{\phi}{1} )
(-\lambda^{-2} -\Omega\lambda^{-3}\met{\lambda}{1}
 +\frac{1}{2}\Omega \lambda^{-2} q^{ab}\met{q}{1}{_{ab}} ) ],
\label{kg0c}
\end{eqnarray}
where $D_a$ is, as before, the derivative operator on constant-$\Omega$
surfaces induced from $\tilde\nabla_a$. 
Evaluating (\ref{kg0c}) and its first two normal derivatives
on ${\cal H}$, we obtain, respectively,
\begin{eqnarray}
0&=& (D^2-1)\met{\phi}{0}, \label{kgcst0}\\
0&=& D^2\met{\phi}{1}, \label{eqphi1}\label{kgcst1}\\ 
0&=& D^2\metbar{\phi}{2}+3\metbar{\phi}{2}
-\met{q}{2}{^{ab}}D_aD_b\met{\phi}{0} - 16\met{\lambda}{1}\met{\phi}{1}
\nonumber \\
&&\ 
-14\met{\lambda}{1} D^a\met{\lambda}{1} D_a\met{\phi}{0}
-2(D_aD_b\met{\lambda}{1}) D^a\met{\lambda}{1}D^b\met{\phi}{0}
+32\met{\lambda}{1} {^2} \met{\phi}{0}
+2(D\met{\lambda}{1} ){^2} \met{\phi}{0}. \label{kgcst2}
\end{eqnarray}
For the $n$-th derivative, the equation that results has the form
\begin{equation}
0= (D^2 + n^2 -1)\met{\phi}{n} - 4n^2 (n-1)\ldb\met{\phi}{n-1} + ... ,
\label{kg-gen}
\end{equation}
where $...$ involves only remnants of $\phi$ of order $\le n-2$.

Next, let $\tilde F_{ab}$ be a Maxwell field
asymptotically regular of order $2$.
Then Maxwell's equations yield 
\begin{equation}
0=\tilde \nabla^m \tilde F_{ma}=
-\lambda \Omega D^m E_m\nabla_a\Omega
-\lambda^{-1} \Omega^2 \epsilon_{a}{^{bc}}[
D_b (\lambda B_c )
+\frac{1}{2}\Omega {\cal L}_{\lambda^2 n}( E^m\epsilon_{mbc})].
\label{max0c}
\end{equation}
Evaluating (\ref{max0c}) and its first two normal derivatives
on ${\cal H}$, we obtain, respectively,
\begin{xalignat}{2}
D_a \met{E}{0}{^a}&=0, & D_a 
\met{B}{0}{^a}&=0, 
\label{em0.1} \\
D_{[a} \met{E}{0}_{b]} &= 0, &
D_{[a} \met{B}{0}_{b]} &= 0, 
\label{em0.2}
\end{xalignat}
\bea
D_{[a} \met{{\cal B}}{1}{_{b]}} 
&=&
-\frac{1}{2} \epsilon_{ab}{^{c}}
 (\met{{\cal E}}{1}{_{c}} - 2 \ldb\met{E}{0}{_{c}}),
\label{eme1a}\\
D_{[a} \met{{\cal E}}{1}{_{b]}} 
&=&
\frac{1}{2} \epsilon_{ab}{^{c}}
 (\met{{\cal B}}{1}{_{c}} - 2 \ldb\met{B}{0}{_{c}}),
\label{emb1a}
\eea
\bea
D_{[a} \met{{\cal B}}{2}{_{b]}} 
&=&
- \epsilon_{ab}{^{c}}
 [\met{{\cal E}}{2}{_{c}} -4\ldb\met{{\cal E}}{1}{_{c}}
 +w_{cd}\met{E}{0}{^d} ],
\label{eme2a}\\
D_{[a} \met{{\cal E}}{2}{_{b]}} 
&=&
\epsilon_{ab}{^{c}}
 [\met{{\cal B}}{2}{_{c}} -4\ldb\met{{\cal B}}{1}{_{c}}
 +w_{cd}\met{B}{0}{^d} ],
\label{emb2a}
\eea
where we have set
$\met{{\cal E}}{1}{_{a }} =\met{E}{1}{_{a }} +\ldb\met{E}{0}{_{a }}$,
$\met{{\cal E}}{2}{_{a }} = \met{E}{2}{_{a }} + 2 \ldb \met{E}{1}{_{a }} +
\met{\lambda}{2} \met{E}{0}{_{a }}$,
$\met{{\cal B}}{1}{_{a }} =\met{B}{1}{_{a }} +\ldb\met{B}{0}{_{a }}$,
$\met{{\cal B}}{2}{_{a }} = \met{B}{2}{_{a }} + 2 \ldb \met{B}{1}{_{a }} +
\met{\lambda}{2} \met{B}{0}{_{a }}$, and
$w_{ab} =  -\met{q}{2}{_{ab}}
+ (\frac{1}{2}\met{q}{2} + 3\ldb{^2} -\met{\lambda}{2} ) \met{q}{0}{_{ab}}$. 
For the $n$-th derivative, the equations that result have the form
\bea
D_{[a} (\met{B}{n}{_{b]}} +n \ldb\met{B}{n-1}{_{b]}})
&=&
-\frac{n}{2} \epsilon_{ab}{^{c}}
 (\met{E}{n}{_{c}} -n\ldb\met{E}{n-1}{_{c}} + ... ),
\label{max-gen1}
\\
D_{[a} (\met{E}{n}{_{b]}} +n\ldb\met{E}{n-1}{_{b]}})
&=&
\frac{n}{2} \epsilon_{ab}{^{c}}
 (\met{B}{n}{_{c}} -n\ldb\met{B}{n-1}{_{c}} + ... ),
\label{max-gen2}
\eea
where $...$ involves only remnants of $E_a$, $B_a$ of order $\le n-2$.

We turn finally to the gravitational field.
The remnant field equations of order $\le 2$ 
were obtained by Beig and Schmidt$^{8, 9}$
in the vacuum case under the gauge conditions (\ref{gauge-fixing}).
We here drop the assumption of vacuum and the gauge condition.
See Appendix C for outline of our derivation.
The zeroth-order equations are
satisfied identically.
The first-order equations are
\begin{eqnarray}
\met{q}{1}{_{ab}} &= &-2\ldb\met{q}{0}{_{ab}}
\label{q1eq} \\
(D^2+3)\ldb &=&0.
\label{lameq}
\end{eqnarray}
The second-order equations are 
\bea
\met{q}{2}&=&2(D\ldb )^2+24\ldb {^2}
 -D^2\met{\lambda}{2} -6\met{\lambda}{2} 
-D^m\met{T}{0}{_m} - 2\met{T}{0}
\label{ee32a.1} \\
D_b\met{q}{2}_a {^{\ b}}&=&32\ldb D_a \ldb +4D^b\ldb D_aD_b\ldb 
 -4D_a\met{\lambda}{2} -6D_a (D^2\met{\lambda}{2} ) \nonumber \\ &&\ \ \ \
+D_a (-D^m\met{T}{0}{_m} - 2\met{T}{0})
+2\met{T}{0}{_a},
\label{ee32a.2} \\
(D^2-2)\met{q}{2}_{ab}&=&
8(D\ldb )^2\met{q}{0}{_{ab}}
+20 D_a\ldb D_b\ldb
+28\ldb D_aD_b\ldb  
-36\ldb {^2}\met{q}{0}{_{ab}}
+4D_aD^c\ldb D_bD_c\ldb
\nonumber   \\ &&\ \ \ 
+4D^c\ldb D_aD_bD_c\ldb 
 -4D_aD_b\met{\lambda}{2} +4\met{\lambda}{2}\met{q}{0}{_{ab}}
 -D_aD_b(D^2\met{\lambda}{2}) \nonumber   \\ &&\ \ \ 
-4\met{T}{0}{_{ab}}
+ D_aD_b (-D^m\met{T}{0}{_m} - 2\met{T}{0})
+4D_{(a}\met{T}{0}{_{b)}}.
\label{ee32a.3} 
\eea
The third- and fourth-order equations are not used in what follows and
are collected in Appendix C, where we also rewrite the second-order
equations in terms of the remnants of the Weyl tensor.

We turn now to the issue of whether or not
the various conserved quantities that we defined in
flat space-time can be generalized to curved space-time.
Recall that a conserved quantity is given by the integral
over a cut of ${\cal H}$ of a vector field $v^a{_\Gamma}$
on ${\cal H}$, where that field is expressed as an
algebraic function of the preferred field $\alpha_\mu$
of the universal background geometry of ${\cal H}$,
the remnants of the physical field, the remnants of the
geometry, and their derivatives. The divergence
of this $v^a{_\Gamma}$ must,
for independence of cut, vanish by virtue of the equations
satisfied by $\alpha_\mu$ and the  various remnants.
In the special case of flat space-time, we have
(or, at least, achieved via gauge)
$\met{\lambda}{k} = 0$,
$\met{q}{k}{_{ab}} = 0$, for $k\ge 1$,
i.e., we have effectively no ``remnants of the geometry''.
Clearly, every conserved quantity in general remains a
conserved quantity in the special case of flat space-time.
But the converse need not hold. Given a conserved quantity
in flat space-time---i.e., given a divergence-free field
$v^a{_\Gamma}$ constructed from the preferred field $\alpha_\mu$,
the remnants of the physical field, and their 
derivatives---then it may or may not be the case that it is the
specialization to flat space-time of some conserved quantity in
curved space-time. When it is, we say we have produced a 
{\em generalization} of our given flat-space conserved quantity.

Consider first the Klein-Gordon case.
We have immediately from eqn.(\ref{eqphi1}):
\begin{theorem}
The conserved quantity ${\cal K}$ ($n=1$ in eqn.(\ref{K}))
for the Klein-Gordon field
in flat space-time admits a generalization,
in the sense described above, to a conserved
quantity in curved space-time,
namely that given by
\begin{equation}
{\cal K} = \int D^a\met{\phi}{1} dS_a .
\end{equation}
\end{theorem}

For the higher-order Klein-Gordon remnant radiation multipoles, we have
\begin{theorem}
Fix $n\ge 2$. 
Then the conserved quantity
${\cal K}{_{\mu_1\cdots\mu_{n-1}}}$ 
for the Klein-Gordon field in flat space-time, 
given by eqn.(\ref{K}), 
does not admit generalization to curved space-time.
\label{thm5}
\end{theorem}
{\em Proof}\/: 
Let, for contradiction, 
$v^a {_{\mu_1\cdots\mu_{n-1}}} $ be a generalization
to curved space-time.
By a simple scaling argument (using, respectively,
linearity of the Klein-Gordon
remnant field equations in the $\met{\phi}{k}$
and homogeneity of all remnant field equations in order), we may assume
that $v^a{_{\mu_1\cdots\mu_{n-1}}}$ is linear in 
the $\met{\phi}{k}$, and
of total order $n$ in all remnants.
Thus, $v^a{_{\mu_1\cdots\mu_{n-1}}}$ contains no $\met{\phi}{k}$, for $k > n$,
and the term involving $\met{\phi}{n}$ is, 
because 
$v^a {_{\mu_1\cdots\mu_{n-1}}} $ 
must reduce to 
the integrand of 
${\cal K} {_{\mu_1\cdots\mu_{n-1}}} $ 
in flat space-time,
precisely
$\psi_{\mu_1\cdots\mu_{n-1}} D^a \met{\phi}{n}
- \met{\phi}{n} D^a\psi_{\mu_1\cdots\mu_{n-1}} $,
where we have set 
$ \psi_{\mu_1\cdots\mu_{n-1}} 
={\cal C}( \alpha_{\mu_1}\cdots\alpha_{\mu_{n-1}})$.
Denote by $u^a{_{\mu_1\cdots\mu_{n-1}}}$ the term of $v^a{_{\mu_1\cdots\mu_{n-1}}}$ involving $\met{\phi}{n-1}$. 
Then, the vanishing of divergence of
$v^a {_{\mu_1\cdots\mu_{n-1}}} $ 
implies
\begin{equation}
D_a u^a{_{\mu_1\cdots\mu_{n-1}}} \hat{=} -4n^2 (n -1)\ldb\psi{_{\mu_1\cdots\mu_{n-1}}}\met{\phi}{n-1} ,
\label{ueq}
\end{equation}
where $\hat{=}$ stands for equality modulo 
Klein-Gordon remnants of order $\le n-2$.
But there exists no such
$u^a{_{\mu_1\cdots\mu_{n-1}}}$, as one sees by the following steps. 
First,
add to $v^a{_{\mu_1\cdots\mu_{n-1}}}$ a divergence of
an antisymmetric tensor field to achieve
the form
\begin{equation}
u^a{_{\mu_1\cdots\mu_{n-1}}} 
= w{_{\mu_1\cdots\mu_{n-1}}} D^a \met{\phi}{n-1} 
- \met{\phi}{n-1} D^a w{_{\mu_1\cdots\mu_{n-1}}},
\end{equation}
with $w{_{\mu_1\cdots\mu_{n-1}}}$ 
linear in $\ldb$ and $\psi{_{\mu_1\cdots\mu_{n-1}}}$, 
and from eqn.(\ref{ueq}) and (\ref{kg-gen}) satisfying
\begin{equation}
(D^2  + n^2 - 2n)w
{_{\mu_1\cdots\mu_{n-1}}}= 4n^2 (n -1)\ldb \psi_{\mu_1\cdots\mu_{n-1}} .
\label{weq}
\end{equation}
Second, 
replace every occurrence of \ldb\ in eqn.(\ref{weq})
by $\alpha_\mu$. Then, under this substitution, 
$w{_{\mu_1\cdots\mu_{n-1}}}$ reduces to the form
$w{_{\mu_1\cdots\mu_{n-1}}} = c D_a\alpha_\mu D^a\psi_{\mu_1\cdots\mu_{n-1}} 
+ c' \alpha_\mu \psi_{\mu_1\cdots\mu_{n-1}} $,
for some constants $c$, $c'$.
Since $\alpha_\mu$ satisfies
eqn.(\ref{lameq}), which is the only property of \ldb\ that
may be used in establishing (\ref{weq}), 
it follows that
eqn.(\ref{weq}) must continue to hold after replacing
\ldb\ therein by $\alpha_\mu$. 
However, under this substitution, eqn.(\ref{weq}) becomes
\begin{equation}
2 [(n+1)c + c'] [
D_a\alpha_\mu D^a\psi_{\mu_1\cdots\mu_{n-1}} 
+ (n -1)\alpha_\mu \psi_{\mu_1\cdots\mu_{n-1}} ]
= 4n^2 (n -1)\alpha_\mu \psi_{\mu_1\cdots\mu_{n-1}} ,
\label{weq2}
\end{equation}
which can never hold.
\ \ $\Box$

We turn next to Maxwell fields.  We have
\begin{theorem}
Let $\met{B}{0}{_a} = 0$, and let the 
the stress-energy tensor $T_{ab}$ satisfy 
\begin{equation}
\met{T}{0}{_a}  = 0,
\hspace{2cm}
\met{T}{0}{_{ab}}\met{E}{0}{^b} - \met{T}{0}\met{E}{0}{_a} 
= 0.
\label{tcond}
\end{equation}
Then the conserved quantity ${\cal E}_{\mu\nu}$
(of eqn.(\ref{caledef})) 
for the Maxwell field
in flat space-time admits a
generalization to a conserved quantity in curved space-time,
namely that given by
\begin{equation}
{\cal E}_{\mu\nu} = 
\int s^{ab} \alpha_{[\nu} D_{|b|} \alpha_{\mu ]} dS_a,
\label{cemunu}
\end{equation}
where we have set
\begin{eqnarray}
s_{ab} &=& 
2 D_{(a} \met{\cal E}{2}{_{b)}} -2 D^c \met{\cal E}{2}{_{c}} \met{q}{0}{_{ab}}
+ 16 \met{\cal E}{1}{_{(a}} D_{b)} \met{\lambda}{1}  
-8 \met{\cal E}{1}{^{c}} D_c\met{\lambda}{1}  \met{q}{0}{_{ab}}
+ 16\psi\met{\lambda}{1}
D_aD_b\met{\lambda}{1} 
+8\met{\lambda}{1} \met{E}{0}{_{(a}} D_{b)}\met{\lambda}{1}
+8\met{\lambda}{1}{^2}D_a\met{E}{0}{_b}
\nonumber \\ &&\ \ 
+ [12\psi\met{\lambda}{1}{^2} 
-20\met{\lambda}{1}\met{E}{0}{^c}D_c\met{\lambda}{1} 
-4\psi (D\met{\lambda}{1})^2
]\met{q}{0}{_{ab}}
+4\psi w{_{ab}} 
+4\psi \met{T}{0}{_{ab}}
-4 \psi \met{T}{0} \met{q}{0}{_{ab}}
\nonumber \\ &&\ \ 
-11(\psi D_aD_b\met{\lambda}{2}
    +\met{\lambda}{2} D_aD_b\psi - 2 D_{(a}\psi D_{b)}\met{\lambda}{2} )
- \psi ( 11 D^2 \met{\lambda}{2} + 22 \met{\lambda}{2} ) \met{q}{0}{_{ab}},
\label{sab}
\end{eqnarray}
where $\psi$ is so chosen to satisfy
$D_a\psi = \met{E}{0}{_a}$.$^{27}$
\label{thm6}
\end{theorem}
The integrand reduces, in flat space-time, to the integrand
of ${\cal E}_{\mu\nu}$ therein plus a divergence, namely 
$D_b (\met{E}{2}{^{[a}} \alpha_{[\mu} D^{b]}\alpha_{\nu]} )$,
of an antisymmetric tensor field.
The demonstration of $D_b s^{ab} =0$ is given in Appendix B.
The above conditions, (\ref{tcond}), on $T_{ab}$ are satisfied  
when the space-time is vacuum, and also when the source
is the Maxwell field itself. But the condition (\ref{tcond})
need not be satisfied in the presence of other matter sources.
It is readily verified that this generalized ${\cal E}_{\mu\nu}$ 
is again gauge invariant.

%
%

For higher-order Maxwell remnant radiation multipoles, we have
\begin{theorem}
Fix $n\ge 3$. Then the conserved quantity
${\cal E}{_{\mu\mu_1\cdots\mu_{n-1}}}$ 
for the Maxwell field in flat space-time,
given by eqn.(\ref{caledef}),
does not admit generalization to curved space-time.
\label{thm7}
\end{theorem}
The proof of theorem \ref{thm7} is similar to that of
theorem \ref{thm5} and is therefore omitted.

\section{CONCLUSION}

We have constructed, 
for each of a Klein-Gordon field, 
a Maxwell field,
and a linearized gravitational field
in Minkowski space-time, a hierarchy
of conserved quantities which we call 
the remnant radiation multipoles.
In the cases of Klein-Gordon and Maxwell,
we have generalized the remnant radiation monopoles to curved space-time. 
There follows a discussion of some outstanding issues.

Does the remnant gravitational monopole
admit generalization to curved space-time?
We conjecture that the answer is yes. 
In Appendix C we give
the remnant field equations necessary for addressing this
question.
We also there display a candidate for a curved-space
gravitational remnant monopole. 
This candidate has the attractive
feature that its divergence, 
which could in principle have contained remnants or order as high as 3,
contains only remnants of order $\le 2$. 
Although 
the existence of this candidate lends some 
support to the conjecture,
it is of course far from a
proof of it. 
Work is in progress to settle this conjecture.
We further conjecture that none of the higher-order 
remnant radiation multipoles for linearized gravitational fields
admit generalization to curved space-time.

The way we introduced the Klein-Gordon and Maxwell
remnant radiation monopoles in a curved space-time involves 
a quite strong fall-off condition, namely $\met{B}{0}{_{ab}}=0$,
on the gravitation remnants.
In the Klein-Gordon case, this restriction is in fact unnecessary. 
Indeed, in the absence of this condition, the
first-order remnant field equation on $\met{\phi}{1}$ 
becomes $D_av^a = 0$, with 
$ v^a = D^a\met{\phi}{1} - \met{q}{1}{^{ab}}D_b \met{\phi}{0}
+\frac{1}{2}\met{q}{2} D^a\met{\phi}{0}
+\ldb D^a\met{\phi}{0}$.
Thus, ${\cal K}\equiv\int_C v^adS_a$ remain conserved
in asymptotic conditions weaker than the ones presently imposed.
Can other conserved quantities be defined with such weaker asymptotic
conditions?

Do there exist conserved quantities analogous 
to our remnant radiation multipoles, but defined at
null rather than spatial infinity?
And if so, are there any simple relations between the
values of corresponding quantities at spatial and null infinity?
In Minkowski space-time, it should not be too difficult
to answer these two questions.
A relevant observation$^{28}$
is that, in the case of Minkowski space-time 
corresponding conserved quantities in general
take different values at spatial and null infinity.
This result suggests that ``remnant radiation'' is capable of 
escaping between spatial infinity and null infinity.
Recall that 
Newman, Penrose and Exton$^{29,30}$
have introduced certain conserved quantities at null
infinity in curved space-time.
Are these quantities analogs, in any sense, of 
the remnant radiation monopoles?

It is unfortunate that the present treatment of asymptotic
quantities involves such complicated algebra.
It is not entirely clear whether 
these complications are inherent in the subject itself,
or merely a reflection of the present techniques.
One case in which we know that these techniques 
are the culprit
is that of stationary space-times. 
It is not hard to convince oneself that
the present framework, in the case of stationary space-times, is
essentially equivalent to the usual 
formalism involving a 3-dimensional 
manifold of trajectories. 
Since the stationary gravitational 
multipole moments of all order can be defined 
within this 3-dimensional formalism,
it should also be possible,
in principle, to define these very same moments
within the present framework.
However, it already seems difficult to define even the first 
few stationary multipoles in the present framework.
Unlike the 3-dimensional formalism, the present framework is not
well adapted to the 
presence of 
Killing fields. 
For example, to treat Killing's equation order by order yields a
complicated set of remnant equations. 
Finding a more natural way 
of dealing with stationary space-times within the present framework 
may give some clue as to how to tame its algebraic complexity.
Indeed, it may further lead to a
generalization of the stationary multipoles to 
more general asymptotically flat space-times.

We have here restricted our consideration to conserved
quantities that are linear
in (the highest-order part of the remnants of) the physical fields. 
More generally, one might allow polynomial
dependence on the remnants.
A candidate for a conserved quantity
quadratic in the remnants has been given by R. Beig$^{9}$:
$\int \epsilon^{kl(a}D_k\ldb \met{E}{0}{_l}{^{b)}}
\alpha_{[\mu} D_{|b|}\alpha_{\nu ]} dS_a$,
where $\xi^a$ is any Killing field in ${\cal H}$.
However, as shown in Ref. 8, 
this quantity vanishes identically
by virtue of the second-order gravitational remnant equations.
It would be of interest to carry out a systematic 
search for polynomial conserved quantities.
One might even search for conserved quantities with non-polynomial
dependence on the remnant fields,
but the fact that these remnant fields have complicated
gauge behavior rather suggests that no such
quantities will exist.

\section*{ACKNOWLEDGEMENT}
I am very grateful to  R. Geroch, my thesis advisor,
for his guidance, encouragement, and many helpful suggestions.
I wish to thank R. M. Wald for several valuable suggestions.
This work was supported in part by NSF grant
PHY 95-14726 to the University of Chicago.


\appendix

\renewcommand{\thesection}{Appendix \Alph{section}:}
\renewcommand{\thesubsection}{\Alph{section}.\arabic{subsection}}

\section{\!Stationary Fields in Minkowski Space-time}

\setcounter{equation}{0}
\renewcommand{\theequation}{\Alph{section}\arabic{equation}}

Consider in Minkowski space-time, a physical field that is static,
i.e., that it is invariant under a time-translation in the space-time.
In this appendix, we do two things:
Express within the present framework the ordinary static
multipole moments of such a field;
and show that, in this static case, all the remnant radiation
multipoles vanish.
We will only discuss linearized gravity here since the treatment
of Klein-Gordon and Maxwell fields is similar and simpler.

Let $\tilde K_{abcd}$ be a linearized
gravitational field in Minkowski space-time $\tilde M$,
asymptotically regular of order 3.
Further, let $\tilde K_{abcd}$ be static, i.e., let
\begin{equation}
{\cal L}_{\tilde t}\tilde K_{abcd}=0,
\label{k-stationary}
\end{equation}
where $\tilde t^a$ is a (unit) time-like Killing field in
$\tilde M$.
Denote by 
$\zeta$
($=\Omega^{-2}{\cal L}_{\tilde t}\Omega )$ 
the corresponding unit 
time-translation on ${\cal H}$.
Taking the remnants of eqn.(\ref{k-stationary}),
we obtain the following equations on 
the remnant fields$^{31}$
\begin{equation}
{\cal L}_{D\zeta} \met{E}{n}_{ab} 
-(n+1)\zeta\met{E}{n}_{ab}
+2\epsilon^{lm}{_{(a}}\met{B}{n}{_{b)l}}D_m\zeta =0 ,
\label{grstatic}
\end{equation}
for $n = 0, 1, 2, ...$\ .
Set $\met{\phi}{n}_E 
=\met{E}{n}{_{ab}} D^a\zeta D^b\zeta $. 
Then this 
$\met{\phi}{n}_E$ satisfies 
eqn.(\ref{phin}), and, from 
eqn.(\ref{grstatic}),
also
\begin{equation}
{\cal L}_{D\zeta } \met{\phi}{n}{_E} -(n+1)\zeta
\met{\phi}{n}{_E}=0. 
\label{static}
\end{equation}
Under a gauge transformation
$\met{\phi}{n}_E$ changes according to eqn.(\ref{eq2}). 
The same equations hold, similarly, on
$\met{\phi}{n}_B \equiv\met{B}{n}{_{ab}} D^a\zeta D^b\zeta $,
$n = 0, 1,...$ .
We note that
the $\met{E}{n}{_{ab}}$ and $\met{B}{n}{_{ab}}$
for this static linearized gravitational field 
can be expressed in terms of
$\met{\phi}{n}{_E}$ and
$\met{\phi}{n}{_B}$.
Indeed, we have, from
eqn.(\ref{grstatic})
and
eqn.(\ref{grn}),
that 
\begin{eqnarray}
\met{E}{n}{_{ab}} &= &
\frac{
2\zeta {^2} + 1
}{(n+1)(n+2)} 
D_aD_b\met{\phi}{n}{_E}
+
\frac{
3
D_{(a} \zeta D_{b)}D_m\met{\phi}{n}{_E} D^m \zeta
}{2(n+1)(n+2)} 
-
\frac{
D^{k}D^l\met{\phi}{n}{_E} D_{k} \zeta D_l \zeta \met{q}{0}{_{ab}}
}{2(n+1)(n+2)} 
\nonumber \\ &&\ \ 
+
\frac{
5
\zeta D_{(a}\zeta D_{b)}\met{\phi}{n}{_E}
}{2(n+1)} 
+
\frac{
4n + 5
}{2(n+2)} 
\met{\phi}{n}{_E} 
D_a\zeta
D_b\zeta
+
\frac{
(n-3)\zeta {^2} + 2n+1
}{2(n+2)} 
\met{\phi}{n}{_E} \met{q}{0}{_{ab}}
\nonumber \\ &&\ \ 
+
\frac{
2
\zeta \epsilon^{kl}{_{(a}}D_{b)}D_k\met{\phi}{n}{_B} 
D_l \zeta
}{(n+1)(n+2)} 
+
\frac{
2
\epsilon^{kl}{_{(a}}D_{b)} \zeta D_k\met{\phi}{n}{_B} 
D_l \zeta
}{n+1} ,
\label{eab-in-phi}
\end{eqnarray}
and similarly for $\met{B}{n}{_{ab}}$.

Now consider, for 
$n = 0,1,2, ...$
\begin{equation}
M_{\mu_1 ...\mu_n }[\met{\phi}{n}{_E}]
\equiv
\frac{(2n+1)!}{2^{n+1}(n!)^3}\int_C
\left [
\met{\phi}{n}{_E}
(\alpha_{{\mu_1}}
+\langle \alpha_{{\mu_1}},\zeta\rangle \zeta )
\cdots
(\alpha_{{\mu_n}}
+\langle \alpha_{{\mu_n}},\zeta\rangle \zeta )
(1+\zeta{^2})^{-1}
D^a\zeta 
\right ]
dS_a .
\label{sm}
\end{equation}
The integrand 
on the right above is divergence-free,
by eqn.(\ref{static}), 
and so eqn.(\ref{sm})
defines, for each $n$, a  conserved quantity,
$M_{\mu_1 ...\mu_n}[\met{\phi}{n}{_E}]$.
These are precisely the ordinary
static electric multipole moments of the linearized
gravitational field.$^{32}$
They are totally symmetric, and satisfy
\begin{align}
0&=\zeta{^{\mu_1}} M_{\mu_1 \mu_2 \mu_3\cdots \mu_n}, \\
0&=\eta^{\mu_1\mu_2}M_{\mu_1 \mu_2 \mu_3\cdots \mu_n}, \label{eq35}\\
\nabla_\mu M_{\mu_1\cdots \mu_{n}}
&= -(2n-1) h_{\mu (\mu_1}M_{\mu_2\cdots \mu_{n})}
+(n - 1) h_{(\mu_1\mu_2}M_{\mu_3\cdots \mu_{n})\mu}, 
\label{smpos}
\end{align}
where we have set $h_{\mu\nu} = \eta_{\mu\nu} + 
\zeta {_\mu}
\zeta {_\nu}$. 
To prove eqn.(\ref{eq35}), use that 
$\met{\phi}{n}{_E}$
satisfies eqn.(\ref{phin});
To prove eqn.(\ref{smpos}), use the 
gauge behavior, (\ref{eq2}), of $\met{\phi}{n}{_E}$.
Similarly, 
we obtain the magnetic multipole moments,
$M_{\mu_1 ...\mu_n}[\met{\phi}{n}{_B}]$.
These two sets of moments are the
linearized versions of
Hansen's mass and angular momentum
multipole moments, respectively.

Finally, we show that all of the
gravitational remnant radiation multipoles
(the ${\cal G}$'s introduced in 
eqn.(\ref{calgdef}), Section III), 
vanish for a static linearized 
gravitational field in Minkowski space-time.
To see this, substitute 
eqn.(\ref{eab-in-phi})
into the integrand of 
${\cal G}_{\mu\nu\mu_1\cdots\mu_{n-1}}$,
to obtain
\begin{eqnarray}
{\cal G}_{\mu\nu\mu_1\cdots\mu_{n-1}}
&=& {\cal K}[
 c_1 \eta_{0\mu} \eta_{0\nu} \met{\phi}{n}{_E}
+ c_2 \eta_{\mu\nu} \met{\phi}{n}{_E}
+ c_3 \eta_{0(\mu} {\cal L}_{\xi^*_{\nu )}} \met{\phi}{n}{_B}
+ c_4 {\cal L}_{\xi^*_{(\mu}} {\cal L}_{\xi^*_{\nu )}} \met{\phi}{n}{_E}
\nonumber\\
&&\ \ \ \ 
+ c_5 \eta_{0(\mu} {\cal L}_{\xi_{\nu )}} \met{\phi}{n}{_E}
+ c_6 {\cal L}_{\xi_{(\mu}} {\cal L}_{\xi^*_{\nu )}} \met{\phi}{n}{_B}
+ c_7 {\cal L}_{\xi_{(\mu}} {\cal L}_{\xi_{\nu )}} \met{\phi}{n}{_E}
]_{\mu_1\cdots\mu_{n-1}},
\label{eq83}
\end{eqnarray}
where $c_1, ... ,\ c_7$ are certain constants,
and $\xi^a{_\mu} $ and 
$\xi^*{^a}{_\mu} $ are the Killing fields given by
$\xi^a{_\mu} = \zeta D^a \alpha_\mu 
- \alpha_\mu D^a \zeta $ and
$\xi^*{^a}{_\mu} = \epsilon^{abc} D_b\zeta D_c \alpha_\mu $.
Let $C$ denote the $\zeta = 0$ 2-sphere section of  ${\cal H}$.
We show that each term on the right in eqn.(\ref{eq83}) contributes zero
by evaluating the integral over $C$. 
The first four terms contribute zero by
virtue of the fact that 
each of the term satisfies the same equations
as a static $n$-th order Klein-Gordon remnant field $\met{\phi}{n}$,
and that, for any such $\met{\phi}{n}$,
$\int_{C} \met{\phi}{n}\alpha_{\mu_1}\cdots\alpha_{\mu_k} 
D^a\zeta dS_a = 0$, for $0\le k< n$. 
The fifth and sixth terms contribute zero because
for any $\met{\phi}{n}$ as above,
${\cal L}_{\xi_\mu} \met{\phi}{n}$ vanishes on $C$, and
${\cal L}_{D\zeta } ({\cal L}_{\xi_\mu} \met{\phi}{n})$ is a sum
of two terms, one of which 
(namely $(n+1)\zeta {\cal L}_{\xi_\mu} \met{\phi}{n}$)
vanishes on $C$ and the other (namely
$- (\met{\phi}{n}){_\mu} \equiv
- D^a\alpha_\mu D_a\met{\phi}{n}
+ (n+1) \alpha_\mu \met{\phi}{n}$)
satisfies the $(n+1)$th remnant field equation and is static.
Finally the last term contributes zero because 
$({\cal L}_{\xi_\mu} {\cal L}_{\xi_\nu} \met{\phi}{n}{_E})$
is a translation times a term which
satisfies the $(n+1)$th remnant field equation and is static,
and because
${\cal L}_{D\zeta}
({\cal L}_{\xi_\mu} {\cal L}_{\xi_\nu} \met{\phi}{n}{_E})
$ is a sum of two terms, one of which (namely
$(n+1)\zeta 
{\cal L}_{\xi_\mu} {\cal L}_{\xi_\nu} \met{\phi}{n}{_E}
- {\cal L}_{\xi_{(\mu}} (\met{\phi}{n}{_E}){_{\nu )}}$)
vanishes on $C$ and the other (namely 
$ - [{\cal L}_{\xi_{(\mu}} \met{\phi}{n}{_E}]{_{\nu )}}$)
satisfies the $(n+1)$th remnant field equation and 
is equal to a static field $\met{\phi}{n+1}$ on $C$.
\ \ $\Box$

\section{Miscellaneous Results}
\setcounter{equation}{0}
Appendix B.1 contains the proofs of
item $(i)$ of each of theorems 1--3.
Appendix B.2 contains the proofs of 
item $(iii)$ of each of theorems 1--2.
Appendix B.3 completes the proof that 
the ${\cal E}_{\mu\nu}$ we introduced in 
Theorem \ref{thm6}
is indeed conserved.

\subsection{The Remnant Radiation Multipoles
Exhaust the Conserved Quantities in Minkowski Space-time}

We first show that, in the Klein-Gordon case, the ${\cal K}$'s
of eqn.(\ref{K}) exhaust all conserved
quantities in Minkowski space-time linear in $\met{\phi}{n}$ and
multi-linear in ${\cal T}$.
Sketch of proof: 
Let $v^a{_\Gamma}$ be a divergence-free
vector field on ${\cal H}$, constructed linearly in
the $\met{\phi}{n}$, and 
multi-linearly in ${\cal T}$. (We introduce the subscript
$\Gamma$ to stand for any
Greek indices that may be attached to $v^a$.)
Since the various $\met{\phi}{n}$ are uncoupled in (\ref{phin})
we may take $v^a{_\Gamma}$ to depend on just a single
remnant field, say $\met{\phi}{n}$.
Then $v^a{_\Gamma}$ takes the form
\begin{equation}
v^a{_\Gamma}=\sum_{k=0}^{s}w^{aa_1\cdots a_k}{_\Gamma}
D_{a_1} \cdots D_{a_k}\met{\phi}{n},
\label{eqvkg}
\end{equation}
where $s$ is the order of the highest derivative in $v^a{_\Gamma}$.
We may assume
$w^{aa_1\cdots a_s}{_\Gamma} =
w^{(aa_1\cdots a_s )}{_\Gamma} $, since any parts of 
$w^{aa_1\cdots a_s}{_\Gamma} $
antisymmetric between ``$a$'' and another index
can be eliminated 
by adding to $v^a{_\Gamma} $ the divergence of an
antisymmetric second-rank tensor field,
and any parts antisymmetric between two indices neither ``a''
can be eliminated using the definition of
the Riemann tensor.
It now follows, from $D_a v^a{_\Gamma} =0$,
that
$w^{aa_1\cdots a_s }{_\Gamma} = q^{(aa_1}u^{a_2\cdots a_s ) }{_\Gamma} $,
for some tensor field $u^{a_2\cdots a_s }{_\Gamma} $.
Were $s\ge 2$, then this term
could now be eliminated in its entirety by
adding to $v^a{_\Gamma} $ 
a divergence, 
namely $D_{a_2}[2(D^{[a}D_{a_3}\cdots D_{a_s} \met{\phi}{n} )
u^{a_2]a_3\cdots a_s}{_\Gamma} ]$,
of an antisymmetric tensor.
So we may assume $s=1$ in eqn.(\ref{eqvkg}),
i.e., we may set
$v^a= w{_\Gamma} D^a\met{\phi}{n}- \met{\phi}{n} w^a{_\Gamma} $. 
It now follows, 
again from $D_a v^a{_\Gamma} =0$, 
that $w^a{_\Gamma} =D^aw{_\Gamma} $,
where $w{_\Gamma} $ is some solution of eqn.(\ref{phin}).
But this $w{_\Gamma} $ must be multi-linear in ${\cal T}$,
and the only$^{33}$
such solution of eqn.(\ref{phin})
is ${\cal C}( \alpha_{\mu_1}\cdots \alpha_{\mu_{n-1}})$.$^{34}$

We next show that, in the Maxwell case, the ${\cal E}$'s 
of eqn.(\ref{caledef}) 
together
with the electric and magnetic charges
exhaust all conserved quantities 
in Minkowski space-time linear in
the remnants of the Maxwell field and multi-linear in ${\cal T}$. 
Sketch of proof:
Let $v^a{_\Gamma}$ be a divergence-free vector field
on ${\cal H}$, constructed linearly
in the $\met{E}{n}{_a}$ and multi-linearly in ${\cal T}$.
As before, we may take $v^a{_\Gamma}$ to depend on a single remnant
field, $\met{E}{n}{_a}$. 
If $n=0$, the result, that 
$v^a = \met{E}{0}{^a}$, 
follows by setting
$\met{E}{0}{_a} = D_a\psi$ with $D^2\psi = 0$, and
using the Klein-Gordon result.
So, let $n\ge 1$.
Then $v^a{_\Gamma}$ takes the form
\begin{equation}
v^a{_\Gamma}=\sum_{k=0}^{s}w^{aa_1\cdots a_{k+1}}{_\Gamma}
D_{a_1} \cdots D_{a_k}\met{E}{n}{_{a_{k+1}}},\ \ \ \ \ n\ge 1,
\label{vmax}
\end{equation}
where $s$ is the order of the highest derivative in $v^a{_\Gamma}$.
An argument similar to the Klein-Gordon case shows
that $v^a{_\Gamma}$ can be brought to the form
\begin{equation}
v^a{_\Gamma} = w_b{_\Gamma} D^a\met{E}{n}{^b} 
-\met{E}{n}{_b} D^a w^b{_\Gamma}+\mu{_\Gamma} \met{E}{n}{^a},
\end{equation}
where $(D^2+n^2 -2) w_a {_\Gamma}=D_a\mu{_\Gamma} $.
We may achieve
$\mu{_\Gamma}  = 0$ in (\ref{vmax})
by adding to $v^a{_\Gamma}$ a divergence
of an antisymmetric tensor field,
namely
$D_b (2\met{E}{n}{^{[a}}D^{b]}w{_\Gamma}
+2w{_\Gamma}D^{[a}\met{E}{n}{^{b]}} 
+\frac{2}{n^2} c{_\Gamma}D^{[a}\met{E}{n}{^{b]}})$
where $c_\Gamma$ is a certain constant and
where we have set
$w{_\Gamma} = \frac{1}{n^2}(-\mu{_\Gamma} + D_a w^a{_\Gamma})$.
Now $w^a{_\Gamma}$ satisfies precisely the same
equations as $\met{E}{n}{^a}$.
The conserved quantity thus arises
from the ``symplectic product'' between $\met{E}{n}{_a}$ and $w_a{_\Gamma}$.
But this $w^a{_\Gamma}$ must be multi-linear in ${\cal T}$,
and the only such solution of 
eq.(\ref{maxn}) 
is 
$w^a{_{\mu\mu_1\cdots\mu_{n-1}}}
= 
{\cal C} (\alpha_\mu \cdots \alpha_{\mu_{n-1}} )
D^a \alpha_\mu
+\frac{1}{n^2} D^a ( 
D^b\alpha_\mu D_b {\cal C} (\alpha_\mu \cdots \alpha_{\mu_{n-1}} )
- \alpha_\mu  {\cal C} (\alpha_\mu \cdots \alpha_{\mu_{n-1}} )
$.

Finally, we show that, in the case of linearized gravity,
the ${\cal G}$'s (of eqn.(\ref{calgdef})) together
with the mass-momentum and angular momentum 
exhaust all conserved quantities in Minkowski space-time
linear in remnants of the linearized gravitational field
and multi-linear in ${\cal T}$. 
Sketch of proof:
Let $v^a{_\Gamma}$ be a divergence-free vector field
on ${\cal H}$, constructed linearly
in the $\met{E}{n}{_{ab}}$ and multi-linearly in ${\cal T}$.
As before, we may take $v^a{_\Gamma}$ to depend on a single remnant
field, $\met{E}{n}{_{ab}}$. 
If $n=0$, the result, that 
$v^a{_\mu} = \met{E}{0}{^{ab}}D_b\alpha_\mu $, 
follows by setting
$\met{E}{0}{_{ab}}=D_aD_b\psi +\psi \met{q}{0}{_{ab}}$
with $D^2\psi = -3\psi$, and using the Klein-Gordon result.
If $n=1$, the result, that 
$v^a{_{\mu\nu}} = \met{E}{1}{^{ab}}\alpha_{[\mu} D_b\alpha_{\nu ]}$, 
follows by setting
$\met{E}{1}{_{ab}}=D_{(a} u_{b)}$
with $D^2 u_a= -2u_a$, $D_au^a = 0$, 
and using the Maxwell result.
So, let $n\ge 2$.
Then $v^a{_\Gamma}$ takes the form
\begin{equation}
v^a{_\Gamma}=\sum_{k=0}^{s}w^{aa_1\cdots a_{k+2}}{_\Gamma}
D_{a_1} \cdots D_{a_k}\met{E}{n}{_{a_{k+1}a_{k+2}}},\ \ \ \ \ n\ge 2,
\end{equation}
where $s$ is the order of the highest derivative in $v^a{_\Gamma}$.
An argument similar to the Klein-Gordon case shows that
$v^a{_\Gamma}$ can be brought to the form
\begin{equation}
v^a{_\Gamma} = w_{bc}{_\Gamma} D^a\met{E}{n}{^{bc}} 
-\met{E}{n}{_{bc}} D^a w^{bc}{_\Gamma}+\met{E}{n}{^{ab}}u_b{_\Gamma} ,
\label{vgl} 
\end{equation}
where $w_{ab}{_\Gamma}$ is symmetric and trace-free, and
satisfies
$(D^2+n^2-3)w_{ab}{_\Gamma} =D_{(a} u_{b)}{_\Gamma}-\frac{1}{3}
\met{q}{0}{_{ab}}
D_mu^m{_\Gamma} $.
We may achieve $u_a {_\Gamma}= 0$ in (\ref{vgl}) by
adding to $v^a{_\Gamma}$ a divergence of an antisymmetric tensor field.
The result now follows from an argument similar
to the Maxwell case.

\subsection{Gauge Behavior of Remnant Radiation Multipoles}

We here prove that the gauge behavior of the Klein-Gordon and Maxwell
remnant radiation multipoles is that given respectively by 
eqn.s (\ref{mult}) and (\ref{multe}).

For the Klein-Gordon case, denote by
${k}{^a}{_{\mu_1\cdots\mu_{n-1}}}$
the integrand of eqn.(\ref{K}).
Then we have
\begin{align}
&\hspace{0.5cm}
\nabla_{\mu} 
{k}{^a}{_{\mu_1\cdots\mu_{n-1}}}
\nonumber \\
&= n 
D^a 
(
{\cal L}_{D\alpha_{\mu}} 
\metbar{\phi}{n-1}
-n\alpha_\mu \metbar{\phi}{n-1}
 )
{\cal C}(\alpha_{\mu_1} \cdots\alpha_{\mu_{n-1}})
-n 
(
{\cal L}_{D\alpha_{\mu}} 
\metbar{\phi}{n-1}
-n\alpha_\mu \metbar{\phi}{n-1}
 )
D^a {\cal C}(\alpha_{\mu_1} \cdots\alpha_{\mu_{n-1}}) .
\nonumber \\
&=
\frac{1}{2}n(n-2)\eta_{(\mu_1\mu_2}
{k}^a_{\mu_3\cdots\mu_{n-1} )\mu}-
n(n-1)\eta_{\mu (\mu_1}
{k}^a_{\mu_2\cdots\mu_{n-1} )}
\nonumber \\
&\hspace{0.5cm} +D_m 
\left [
2nD^{[a}\met{\phi}{n-1} D^{m]}\alpha_\mu
{\cal C}(\alpha_{\mu_1} \cdots\alpha_{\mu_{n-1}})
+ 2n \met{\phi}{n-1} 
D^{[a} \alpha_\mu
D^{m]}
{\cal C}(\alpha_{\mu_1} \cdots\alpha_{\mu_{n-1}})
\right ].
\end{align}
where we used, in the first step, 
eqn.s (\ref{K}) and (\ref{eq2}), 
and, in the second,
the identity$^{35}$
\begin{eqnarray}
D^m\alpha_\mu D_m
{\cal C}( \alpha_{\mu_1} \cdots \alpha_{\mu_{n-1}})
&=&-(n-1)
\alpha_\mu {\cal C}( \alpha_{\mu_1} \cdots \alpha_{\mu_{n-1}})
+(n-1)\eta_{\mu (\mu_1}
{\cal C}( \alpha_{\mu_2} \cdots \alpha_{\mu_{n-1})})
\nonumber \\
&&\ \ \ \ -\frac{n-2}{2}
\eta_{(\mu_1 \mu_2}
{\cal C}( \alpha_{\mu_3} \cdots \alpha_{\mu_{n-1})}
\alpha_\mu ) ,
\label{id1}
\end{eqnarray}
Integrate over a cut of ${\cal H}$.

For the Maxwell case, denote by 
$({e^E})^a_{\nu\mu_1\cdots\mu_{n-2} }$
the integrand of eqn.(\ref{caledef}),
and set $\xi^a{_{\mu\nu}} = 2\alpha_{[\mu } D^a \alpha_{\nu ]}$.
Then we have
\begin{eqnarray}
\nabla^{\mu} [({e^E})^a ]^\nu {_{\mu_1\cdots\mu_{n-1} }}
&=& n 
D^a 
\left [
({\cal L}_{D\alpha_{\mu}} 
\met{E}{n-1}{_m}
-n\alpha_\mu \metbar{E}{n-1}{_m}
+\epsilon_{mkl}\met{B}{n-1}{^k}D^l\alpha_\mu  )D^m\alpha_\nu 
\right ]
{\cal C}(\alpha_{\mu_1} \cdots\alpha_{\mu_{n-1}})
\nonumber \\
&&-n 
\left [
({\cal L}_{D\alpha_{\mu}} \met{E}{n-1}{_m}
-n\alpha_\mu \metbar{E}{n-1}{_m}
+\epsilon_{mkl}\met{B}{n-1}{^k}D^l\alpha_\mu  )D^m\alpha_\nu 
\right ]
D^a {\cal C}(\alpha_{\mu_1} \cdots\alpha_{\mu_{n-1}})
\nonumber \\
&=&
\nabla_{\mu} {k}^a_{\mu_1\cdots\mu_{n-1} }[\met{E}{n-1}{_m}D^m\alpha_\nu ]
\nonumber \\
&&+\frac{n(n-2)}{n-1}\eta_{(\mu_1\mu_2}
[{(e^E)^a}]
^{[\nu\mu ]}{_{\mu_3\cdots\mu_{n-1} )}}
- 2n\delta^{[\mu }{_{ (\mu_1}}
[{(e^E)^a}]^{\nu ]}{_{ \mu_2\cdots\mu_{n-1} )}}
\nonumber \\ 
&& -\frac{n}{n-1}
D_m 
\left [
2(D^{[a}\met{E}{n-1}{_k}D^{m]}\xi^k{_{\mu\nu}}
+\met{E}{n-1}{^{[a}}\xi^{m]} {_{\mu\nu}})
{\cal C}(\alpha_{\mu_1} \cdots\alpha_{\mu_{n-1}})
\right .
\nonumber \\ &&\ 
\ \ \ \ \ \ \ \ \quad \quad \quad
\left . +2\met{E}{n-1}{^k}D^{[a}\xi_k {_{\mu\nu}}D^{m]}
{\cal C}(\alpha_{\mu_1} \cdots\alpha_{\mu_{n-1}}) 
\right ],
\label{multecal}
\end{eqnarray}
where we used, in the first step,
eqn.s (\ref{eg}) and (\ref{caledef}), and, in the second,
the identity eqn.(\ref{id1}) again.
Integrate over a cut of ${\cal H}$.

\subsection{Completion of Proof of Theorem \ref{thm6}}

In our proof of Theorem \ref{thm6}, we omitted one step: The
demonstration the the $s_{ab}$ of eqn.(\ref{sab})
is indeed divergence-free. We here supply that step.
We have
\begin{align}
D^b s_{ab} 
&= 
- 2 \epsilon_{abc} D^b(w^{cd}\met{B}{0}{_d})
+16\met{\lambda}{1}\epsilon_{abc} D^b\ldb \met{B}{0}{^c}
-8\psi \met{T}{0}{_a}
+ 4(\met{T}{0}{_{ab}}\met{E}{0}{^b} - \met{T}{0}\met{E}{0}{_a} )
\nonumber \\
&= 0,
\end{align}
where, in the first step, we used
the following six equations
\begin{align}
&D^b
(2 D_{(a} \met{\cal E}{2}{_{b)}} -2 D^c \met{\cal E}{2}{_{c}} \met{q}{0}{_{ab}})
= 
16 (D_{[a}\met{{\cal E}}{1}{_{b]}})D^b\met{\lambda}{1}
+ 24 \met{\lambda}{1} \met{{\cal E}}{1}{_{a}}
-16\met{\lambda}{1}{^2}\met{E}{0}{_a}
-4w{_{ab}}\met{E}{0}{^b} 
- 2 \epsilon_{abc} D^b(w^{cd}\met{B}{0}{_d}) 
\nonumber \\
& \hspace{4.8cm}+16\met{\lambda}{1}\epsilon_{abc} D^b\ldb \met{B}{0}{^c} ,
\label{mid1}
\end{align}
\begin{align}
&D^b ( 16 \met{\cal E}{1}{_{(a}} D_{b)} \met{\lambda}{1}  
-8 \met{\cal E}{1}{^{c}} D_c\met{\lambda}{1}  \met{q}{0}{_{ab}})
=
-16 (D_{[a}\met{{\cal E}}{1}{_{b]}})D^b\met{\lambda}{1}
- 24 \met{\lambda}{1} \met{{\cal E}}{1}{_{a}}
+16(\met{E}{0}{^c}D_c\met{\lambda}{1})D_a\met{\lambda}{1} ,
\end{align}
\begin{align}
&D^b ( 4\psi w{_{ab}} ) =
4w {_{ab}} \met{E}{0}{^b} 
- 8\psi\ldb D_a \ldb
- 8\psi D_aD_b\ldb D^b \ldb
+ 22\psi D_a (D^2\met{\lambda}{2})
+\psi D_a (2D_b T^b + 4T) - 8\psi T_a ,
\end{align}
\begin{align}
&D^b [
 16\psi\met{\lambda}{1}
D_aD_b\met{\lambda}{1} 
+8\met{\lambda}{1} \met{E}{0}{_{(a}} D_{b)}\met{\lambda}{1}
+8\met{\lambda}{1}{^2}D_a\met{E}{0}{_b}
+ (12\psi\met{\lambda}{1}{^2} 
-20\met{\lambda}{1}\met{E}{0}{^c}D_c\met{\lambda}{1} 
- 4\psi (D\ldb )^2
)\met{q}{0}{_{ab}}
\nonumber \\ &\quad \ \ 
= 16\ldb\met{E}{0}{_a} 
- 16(\met{E}{0}{^b} D_b\ldb ) D_a\ldb
+8\psi\ldb D_a\ldb
+ 8\psi D_aD_b\ldb D^b\ldb ,
\end{align}
\begin{align}
&D^b [
-11(\psi D_aD_b\met{\lambda}{2}
    +\met{\lambda}{2} D_aD_b\psi - 2 D_{(a}\psi D_{b)}\met{\lambda}{2} )
+ \psi (-11 D^2 \met{\lambda}{2} + 22 \met{\lambda}{2} ) \met{q}{0}{_{ab}},
] = - 22 \psi D_a ( D^2 \met{\lambda}{2}) ,
\end{align}
\begin{align}
&D^b (4\psi \met{T}{0}{_{ab}} - 4\psi \met{T}{0} \met{q}{0}{_{ab}}) =
4(\met{T}{0}{_{ab}} \met{E}{0}{^b} - \met{T}{0}\met{E}{0}{_a} )
+ 4\psi (D^b \met{T}{0}{_{ab}} - D_a \met{T}{0}) ,
\label{mid5}
\end{align}
(themselves consequences of the remnant field equations 
(\ref{max0c})--(\ref{emb2a}),
(\ref{lameq})--(\ref{ee32a.3}),
(\ref{conseq1}), (\ref{conseq2})),
and, in the second step, 
$\met{B}{0}{_a} = 0$ and
eqn.(\ref{tcond}) of the  theorem.

\section{Gravitational Remnant Equations}
\setcounter{equation}{0}
In C.1 we discuss the issue of generalizing the remnant
radiation multipoles from linearized to full gravitation.
In C.2, we outline the derivation of gravitational remnant
field equations.
In C.3 we present an alternative version of
the second-order gravitational remnant field equations,
involving the remnants of the Weyl tensor.

\subsection{Generalization of Gravitational Remnant Radiation Monopole}
In Section IV, 
we generalized the flat-space Klein-Gordon and Maxwell remnant
radiation mo\-no\-poles to curved space-time. 
However, we
have been unable to
determine whether 
there ex\-ists a similar generalization 
for linearized gravity. 
Here is how far we have gotten.

The remnant equations for gravitation 
were given, up to second order,
in (\ref{lameq})--(\ref{ee32a.3}).
We shall need the next two orders. 
The third-order remnant equations are
\bea
\met{q}{3} &=& \met{q}{2}{^{ab}}D_aD_b\ldb + 2 D^aD^b\ldb D_a\ldb D_b\ldb
+ 12 \ldb (D\ldb)^2 - 24 \ldb{^3}, 
\label{q3eq1} \\
D^b\met{q}{3}{_{ab}} &=& D_a \met{q}{3} + 2\met{q}{2}{_{ab}}D^b\ldb
+ 4 (D\ldb)^2 D_a\ldb + 64\ldb{^2}D_a\ldb, \\
D^2\met{q}{3}{_{ab}} &=& - \met{q}{3}{_{ab}}
+D_aD_b\met{q}{3} -12D\ldb\cdot D\met{q}{2}{_{ab}} + 24\ldb\met{q}{2}{_{ab}}
+ 16 D_{(a}(\met{q}{2}{_{b)m}}D^m\ldb) \nonumber \\
&&\ \ \  \  + 16D_{(a}\ldb D_{b)}D_c\ldb D^c\ldb
-4 (D\ldb)^2 D_aD_b\ldb
+ 160\ldb D_a\ldb D_b\ldb - 52\ldb{^2}D_aD_b\ldb
\nonumber \\
&&\ \ \ \ -[5\met{q}{3} - 12\ldb (D\ldb)^2 + 372\ldb{^3} ]\met{q}{0}{_{ab}} .
\label{q3eq3}
\eea 
The fourth-order remnant equations are
\bea
\met{q}{4} &=& \frac{2}{3}\met{q}{3}{^{ab}}D_aD_b\ldb 
+ \frac{1}{3} D\ldb\cdot D\met{q}{3}
+ \frac{10}{3}\met{q}{2}{^{ab}}D_a\ldb D_b\ldb 
+ 5 \ldb \met{q}{2}{^{ab}}D_aD_b\ldb 
+ 2\met{q}{2}{^{ab}} \met{q}{2}{_{ab}}
\nonumber \\
&&\ \ \ \ -\frac{4}{3} (D\ldb)^2 (D\ldb)^2 
+ \frac{92}{3} \ldb{^2} (D\ldb)^2 
+ 10 \ldb D^aD^b\ldb D_a\ldb D_b\ldb
- 192 \ldb{^4} ,
\label{q4eq1} \\
D^b\met{q}{4}{_{ab}} &=& D_a \met{q}{4} 
+ 4\met{q}{3}{_{ab}}D^b\ldb
+ 2\met{q}{3} D_a\ldb 
+ 3 D^b(\met{q}{2}{_{ac}}\met{q}{2}{^c}{_b})
-\frac{9}{4} D_a (\met{q}{2}{_{bc}}\met{q}{2}{^{bc}})
\nonumber \\
&&\ \ \ \ -\frac{3}{2}\met{q}{2}{_{ab}} D^b\met{q}{2}
- 10\ldb\met{q}{2}{_{ab}}D^b\ldb
+ 76 \ldb (D\ldb)^2 D_a\ldb + 832\ldb{^3}D_a\ldb ,
\label{q4eq2}
\\
D^2\met{q}{4}{_{ab}} &=& - 6\met{q}{4}{_{ab}}
+2 D_{(a}D^m\met{q}{4}{_{b)m}}
+2\met{q}{4} \met{q}{0}{_{ab}}
+16\ldb D_{(a}D^m\met{q}{3}{_{b)m}}
+16 D_{(a}\met{q}{3}{_{b)m}}D^m\ldb
+72 \ldb\met{q}{3}{_{ab}}
\nonumber \\ &&\ \ \ \ 
-20\ldb\met{q}{3} \met{q}{0}{_{ab}}
-8\ldb D^2\met{q}{3}{_{ab}}
-16D\ldb \cdot D\met{q}{3}{_{ab}}
+96\ldb{^2} D_{(a}D^m\met{q}{2}{_{b)m}}
+96 \ldb{^2}\met{q}{2}{_{ab}}
-96 \ldb{^2}\met{q}{2}q{_{ab}}
\nonumber \\ &&\ \ \ \ 
+120 \ldb D_{(a}\met{q}{2}{_{b)m}}D^m\ldb
-48\ldb{^2} D^2\met{q}{2}{_{ab}}
-156 \ldb D\ldb \cdot D\met{q}{2}{_{ab}}
+16 D^cD_{(a}\ldb \met{q}{3}{_{b)c}}
\nonumber \\ &&\ \ \ \ 
+ 16D^c\met{q}{3}{_{c(a}}D_{b)}\ldb
-8 \met{q}{3}{^{cd}}D_cD_d\ldb \met{q}{0}{_{ab}}
-8 D^c \met{q}{3}{_{cd}}D^d\ldb \met{q}{0}{_{ab}}
-12 D_c\met{q}{2}{^{cd}}D_{(a}\met{q}{2}{_{b)d}}
\nonumber \\ &&\ \ \ \ 
-12\met{q}{2}{^{cd}}D_cD_{(a}\met{q}{2}{_{b)d}}
+6 D_c\met{q}{2}{^{cd}}D_{d}\met{q}{2}{_{ab}}
+6\met{q}{2}{^{cd}}D_cD_{d}\met{q}{2}{_{ab}}
+192\ldb D_{(a}\ldb D^c\met{q}{2}{_{b)c}}
\nonumber \\ &&\ \ \ \ 
+192\ldb D^c D_{(a}\ldb \met{q}{2}{_{b)c}}
+192 D_{(a}\ldb \met{q}{2}{_{b)c}} D^c\ldb
-96\ldb D_c\met{q}{2}{^{cd}}D_d\ldb \met{q}{0}{_{ab}}
-96\ldb \met{q}{2}{^{cd}}D_cD_d\ldb \met{q}{0}{_{ab}}
\nonumber \\ &&\ \ \ \ 
-96 \met{q}{2}{^{cd}}D_c\ldb D_d\ldb \met{q}{0}{_{ab}}
+432\ldb {^3}D_aD_b\ldb
+1440\ldb{^2}D_a\ldb D_b\ldb
+1488\ldb{^2}(D\ldb)^2 \met{q}{0}{_{ab}}
\nonumber \\ &&\ \ \ \ 
+1296\ldb{^4}\met{q}{0}{_{ab}}
+48\ldb D\ldb\cdot D\met{q}{2}\met{q}{0}{_{ab}}
-3 D\met{q}{2}\cdot D \met{q}{2}{_{ab}}
+6 D_{(a}\met{q}{2}{_{b)c}}D^c\met{q}{2}
\nonumber \\ &&\ \ \ \ 
+4 D\ldb\cdot D\met{q}{3} \met{q}{0}{_{ab}}
+48 (D\ldb)^2 (D\ldb)^2 \met{q}{0}{_{ab}}
+24\met{q}{2}{_{ac}}\met{q}{2}{^{c}}{_b}
-12\met{q}{2}{_{cd}}\met{q}{2}{^{cd}}\met{q}{0}{_{ab}}
\nonumber \\ &&\ \ \ \ 
-48 (D\ldb)^2\met{q}{2}{_{ab}}
-6 D_c\met{q}{2}{_d}{_{(a}} D^d\met{q}{2}{_{b)}}{^c}
-3 D_a\met{q}{2}{_{cd}} D_b\met{q}{2}{^{cd}}
+6 D_c\met{q}{2}{_d}{_{a}} D^c\met{q}{2}{_{b}}{^d}
-D_aD_b\met{q}{4}
\nonumber \\ &&\ \ \ \ 
-8D_aD_b(\ldb\met{q}{3})
+3D_aD_b(\met{q}{2}{^{cd}}\met{q}{2}{_{cd}})
-48D_aD_b(\ldb{^2}\met{q}{2}) .
\label{q4eq3}
\eea 

We begin by noting that, introducing a potential $h_{ab}$ for the
linearized gravitational field, the integrand of
${\cal G}_{\mu\nu\lambda\sigma}$ in flat space-time
in eqn.(\ref{calgdef})
is a multiple of
$D_{(a}\met{h}{4}{_{bc)}}
\xi^b{_{(\mu |(\lambda }}
\xi^c{_{\sigma ) |\nu )}}$, where 
we have set
$\xi^a{_{\mu\nu}} = 
\alpha_{[\mu }D^a \alpha_{\nu ]} $.
Note also that
$D_{(a}\met{h}{4}{_{bc)}}$
is divergence-free, 
by virtue of the remnant field equations on $\met{h}{k}{_{ab}}$.
%
%
Thus, the problem of
generalizing
to curved space-time
the  ${\cal G}_{\mu\nu\lambda\sigma}$ 
of flat space-time
is equivalent
to that of finding a
third-rank, totally symmetric, divergence-free tensor 
field $s_{abc}$ on ${\cal H}$,
constructed from the gravitational
remnants,
such that 
$s{_{abc}}$
reduces, in the case of linearized gravity, to
$D_{(a}\met{h}{4}{_{bc)}}$.
Consider, in this connection, the candidate
$\hat{s}{_{abc}}$
given by
\bea 
\hat{s}{_{abc}}
&=&
D_{(a}\met{q}{4}{_{bc)}}
\nonumber \\ &&\ \ 
+ (\frac{82}{3} + 4 c) \lambda_{(a} \met{q}{3}{_{bc)}}
- (24 + 4 c) \met{q}{0}{_{(ab}} \met{q}{3}{_{c)d}}\lambda^d
+(\frac{20}{3} + 2 c) \ldb D_{(a} \met{q}{3}{_{bc)}}
\nonumber \\ &&\ \ 
-(\frac{10}{3} + c) \lambda^{d}{_{(ab}} \met{q}{3}{_{c)d}}
+ (\frac{2}{3} + \frac{c}{2}) \met{q}{0}{_{(ab}} \lambda_{c)de} \met{q}{3}{^{de}}
+\frac{4}{3} \lambda^{d}{_{(a}} D_b \met{q}{3}{_{c)d}}
\nonumber \\ &&\ \ 
- (\frac{10}{3} +\frac{c}{2}) \met{q}{0}{_{(ab}} D_{c)} \met{q}{3}{_{de}} \lambda^{de}
+ \frac{8}{3} \met{q}{0}{_{(ab}} D^{d} \met{q}{3}{^e}{_{c)}} \lambda_{de}
+ ( 2 + c) \lambda^{d} D{_{(a}} D_b \met{q}{3}{_{c)d}}
\nonumber \\ &&\ \ 
- ( \frac{8}{3} + c) \lambda^{d} D_d D{_{(a}} \met{q}{3}{_{bc)}}
+ c \lambda_{d(a} D^d \met{q}{3}{_{bc)}},
\label{sabc}
\eea 
where $c$ is any constant, 
and where we have set
$\lambda_a \equiv D_a\ldb$,
$\lambda_{ab}\equiv D_aD_b\ldb +\ldb \met{q}{0}{_{ab}}$ and
$\lambda_{abc}\equiv D_a\lambda_{bc}$. 
This $\hat{s}{_{abc}}$
has all the required properties, except that
its divergence, instead of vanishing, includes
remnants of order not exceeding 2.
The issue, then, is whether one can add to this $\hat{s}{_{abc}}$
terms of order not exceeding two to achieve vanishing divergence.
In any case, the mere
existence of this field $\hat{s}{_{abc}}$ lends
support to the conjecture that ${\cal G}_{\mu\nu\lambda\sigma}$
admits a generalization to curved space-time.  
Work is in progress to settle this conjecture.


\subsection{Derivation of Gravitational Remnant Field Equations}
The Einstein equation
gives rise to certain differential equations on
the gravitational remnants,
$\met{\lambda}{k}{_{ab}}$, $\met{q}{k}{_{ab}}$. 
These equations 
for a vacuum space-time 
were first systematically studied by
Beig and Schmidt$^{8, 9}$.
We have here utilized the non-vacuum equations, of order 
one (\ref{q1eq})--(\ref{lameq}),  
two (\ref{ee32a.1})--(\ref{ee32a.3}),
and vacuum equations of 
order three (\ref{q3eq1})--(\ref{q3eq3}),
and four (\ref{q4eq1})--(\ref{q4eq3}).
We summarize how these were derived.
First write
Einstein's equation in $3+1$-form,
adapted to the surfaces $\Omega =$\,constant:
\bea
\Omega^2 T
&=&-\frac{1}{2}[{\cal R}+p^{mn}p_{mn}-p^2 ],  
\label{ee3.1}  \\
\Omega^2 T_a &=&
D{^m} (p_{am}-pq_{am}), 
\label{ee3.2}  \\
\Omega^2 T_{ab}&=&{\cal R}{_{ab}}+2p_a^{\ m}p_{mb}-pp_{ab}
-\lambda^{-1}{D}{_a}{D}{_b}\lambda 
+\lambda^{-1}p_{ab}
-\Omega {\cal L}_{\lambda n} p_{ab} ,
\label{ee3.3} 
\eea
where ${D}{_a}$ denotes the derivative operator of 
the metric $q_{ab}$ of these surfaces,
${\cal R}{_{ab}}$ its Ricci curvature, 
and $p_{ab}$ the rescaled extrinsic curvature
of these surfaces,
defined by
\begin{equation}
p_{ab}\equiv
\Omega q_a{^k}q_b{^l}\tilde\nabla_k 
(\Omega^{-2}\lambda\tilde\nabla_l\Omega)
=-\lambda^{-1}q_{ab}+\frac{1}{2}\Omega
{\cal L}_{\lambda n} q_{ab}. 
\label{ppot}
\end{equation}
Taking the remnants
of eqn.s (\ref{ee3.1})--(\ref{ee3.3})
through fourth order,
we obtain 
eqs. (\ref{q1eq})--(\ref{lameq}),  
(\ref{lameq})--(\ref{ee32a.3}),
(\ref{q3eq1})--(\ref{q3eq3}),
and (\ref{q4eq1})--(\ref{q4eq3}).

We remark, finally, that the conservation equation of the
stress-energy tensor, $\tilde \nabla^a\tilde T_{ab}=0$,
yields, for the zeroth order remnants 
of $\tilde T_{ab}$, the following equations
\begin{eqnarray}
0&=& D^a\met{T}{0}{_a}
+2\met{T}{0}
+2\met{T}{0}{^m}{_m},
\label{conseq1}
\\
0&=& D^b\met{T}{0}{_{ab}}
-D_a (\met{T}{0} +\met{T}{0}{^m}{_m} ).
\label{conseq2}
\end{eqnarray}

\subsection{Second Order Equations in Weyl Remnants}

We first remark that, for any space-time with completion,
the Weyl tensor is asymptotically regular of order 3.
To see this, rewrite
$2\tilde\nabla_{[a}\tilde\nabla_{b]}n_c
=\tilde R_{abc}{^d}n_d$ 
as
\begin{eqnarray}
E_{ab}&=&\Omega^{-1}
(-\met{\cal R}{q}{_{ab}}-p_a^{\ m}p_{bm} +pp_{ab})
+\Omega[
\frac{1}{2}(T_{ab}-\frac{1}{3}q_{ab}T^m{_{m}})-\frac{2}{3}q_{ab}T 
], \label{epot}\\
B_{ab}&=&\Omega^{-1}\epsilon_{mn(a}\met{D}{q}{^m}p^n{_{b)}},
\label{bpot}
\end{eqnarray}
with 
$E_{ab}$ and $B_{ab}$ given by eqn.(\ref{weylebdef}),
$p_{ab}$ given by eqn.(\ref{ppot}), and
$T_{ab}$ given by eqn.(\ref{teq}).
But, by the conditions in definition \ref{def-1}, 
the right sides are smooth on $M$.

We next remark that the gravitational remnant equations,
(\ref{lameq})--(\ref{ee32a.3}), can be written in terms
of the Weyl remnants,
$\met{E}{k}{_{ab}}$,
$\met{B}{k}{_{ab}}$.
To see this,
first take the zeroth-order remnants of eqn.s (\ref{epot}), (\ref{bpot}) 
above, to obtain 
\begin{eqnarray}
\met{E}{0}{_{ab}}&=&-(D_aD_b\ldb +\ldb\met{q}{0}{_{ab}}),
\label{gr0.2}\\
\met{B}{0}{_{ab}}&=&\epsilon_{kl(a}D^k(
\met{q}{1}{^l}{_{b)}} +2\ldb \met{q}{0}{^l}{_{b)}})=0,
\label{gr0.2b}
\end{eqnarray}
and the first-order remnants, 
to obtain 
\begin{align}
\met{E}{1}_{ab}&=
-\frac{1}{2}\met{q}{2}_{ab}
%
%
+[(D\ldb )^2+5\ldb {^2}]\met{q}{0}{_{ab}}
+\ldb D_aD_b\ldb -2D_a\ldb D_b \ldb 
-\frac{1}{2}\met{T}{0}{_{ab}}-(\frac{2}{3}\met{T}{0}
+\frac{1}{6}\met{T}{0}{^m}{_{m}})\met{q}{0}{_{ab}} ,
\nonumber \\
\label{eqwe1} \\
\met{B}{1}_{ab}
&= \frac{1}{2}\epsilon_{mn(a}D^m
\met{q}{2}{^n}{_{b)}}
%
%
.
\label{eqwb1} 
\end{align}
These Weyl remnants satisfy, 
by virtue of eqn.s (\ref{lameq})--(\ref{ee32a.3}), the equations
\bea
D_{[a}\met{E}{0}{_{b]c}}&=&0.
\label{gr0.1}
\\
D_{[a} \met{{\cal E}}{1}{_{b]c}} &=&
\frac{1}{2}
\epsilon_{ab}^{\ \ m}
[\met{B}{1}{_{mc}} 
+4\epsilon_{kl(m}(D^k\ldb )\met{E}{0}{^l}{_{c)}}
+\frac{1}{2}\epsilon_{mc}{^n}\met{T}{0}{_n} ],
\label{eqe1} 
\\
D_{[a}\met{B}{1}{_{b]c}}&=&-\frac{1}{2}
\epsilon_{ab}^{\ \ m}
[\met{{\cal E}}{1}{_{mc}} 
-2\met{T}{0}{_{mc}}
- \met{T}{0} \met{q}{0}{_{mc}}
+D_{c}\met{T}{0}{_{m}} ],
\label{eqb1} 
\eea
where we have set 
\begin{equation}
\met{\cal E}{1}{_{ab}} = \met{E}{1}{_{ab}} -\met{\lambda}{1}\met{E}{0}{_{ab}}
+\frac{1}{2}\met{T}{0}{_{ab}} + (\frac{1}{6}\met{T}{0}{^m}{_m} 
-\frac{2}{3}\met{T}{0} ) \met{q}{0}{_{ab}}.
\end{equation}

Now fix a space-time
with completion, and define
$\met{E}{0}{_{ab}}$,
$\met{E}{1}{_{ab}}$, and
$\met{B}{1}{_{ab}}$ by
eqn.s (\ref{gr0.2})--(\ref{eqwb1}).
Then 
eqn.s (\ref{lameq})--(\ref{ee32a.3})
are equivalent to the statements that the
$\met{E}{0}{_{ab}}$,
$\met{E}{1}{_{ab}}$, 
$\met{B}{1}{_{ab}}$, so defined are trace-free
and  satisfy (\ref{gr0.1})--(\ref{eqb1}).

\newpage

\noindent \hspcb$^1$ R. Arnowitt, S. Deser and C. W. Misner, 
Phys. Rev. {\bf 117}, 1695, (1960);
Phys. Rev. {\bf 121}, 1566, (1961);
Phys. Rev. {\bf 122}, 997, (1961);
{\em Gravitation, an Introduction to 
Current Research}, ed. L. Witten (New York, Wiley, 1962).

\noindent \hspcb$^2$ R. Geroch, J. Math. Phys. {\bf 13}, 956, (1972).

\noindent \hspcb$^3$ R. Geroch, J. Math. Phys. {\bf 11}, 2580, (1970).

\noindent \hspcb$^4$ A. Ashtekar and R. O. Hansen J. Math. Phys.  {\bf 19}, 1542 (1978).

\noindent \hspcb$^5$ A. Ashtekar and A. Magnon-Ashtekar, J. Math. Phys.
{\bf 20}(5), 793 (1979).

\noindent \hspcb$^6$ P. Sommers, J. Math. Phys. {\bf 19}, 549, (1978).

\noindent \hspcb$^7$ S. Persides, 
J. Math. Phys. {\bf 20}, 1731, (1979),
J. Math. Phys. {\bf 21}, 135, (1980),
J. Math. Phys. {\bf 21}, 142, (1980).

\noindent \hspcb$^8$ R. Beig and B. G. Schmidt, Comm. Math. Phys. {\bf 87}, 65, (1982).

\noindent \hspcb$^9$ R. Beig, Proc. R. Soc. London, A {\bf 391}, 295, (1984).

\noindent \hspca$^{10}$ H. Bondi, A. W. K. Metzner and M. J. G. Van Der Berg, 
Proc. Roy. Soc.  London, A {\bf 269}, 21 (1962).

\noindent \hspca$^{11}$ A. Ashtekar and J. D. Romano, Class. Quantum Grav.
{\bf 9}, 1069--1100, (1992).

\noindent \hspca$^{12}$ Sketch of proof:
Let $(\tilde M,\tilde g_{ab})$
be a stationary asymptotically flat 
vacuum space-time with Killing field $\xi^a$. 
Denote by $\mu$ and $\omega$
the norm and twist of the Killing field respectively.
Let $(\tilde V, \tilde h_{ab})$ denote the
(Riemannian) manifold of orbits of the Killing field, 
$(V,\Lambda )$ 
its completion,
$\Omega_G$ a conformal factor and
$h_{ab}=\Omega_G^2\tilde h_{ab}$ (See Ref.\ 3). 
It follows$^{36}$ that each of
$\Omega_G^{-1/2}\mu^{1/4}(\mu -\mu^{-1}+\mu^{-1}\omega^2)$,
$\Omega_G^{-1/2}\mu^{1/4}(\mu^{-1}\omega )$, and
$\mu +\mu^{-1}+\mu^{-1}\omega^2$ 
is smooth on $V$.
Fix any smooth coordinates $x{^i}$ on $V$ near $\Lambda$ 
such that $h_{ij}|_\Lambda =\delta_{ij}$, and Lie drag
them into $\tilde M$ by $\xi^a$.
Perform an inversion on these coordinates 
to obtain $\tilde x^i$ on $\tilde M$. Pick a $\tau '{_a}$
on $V$ satisfying $D_{[a}\tau '_{b]}=-\frac{1}{2}
\mu^{-3/2}\epsilon_{abc}D^c\omega$
and such that $\tau '_a$ is smooth in $y'$ and  vanishes at $\Lambda$
(See, e.g., the appendix of Ref. 36 for motivation).
Let $\tau_a$ be the pull-back of $\tau'_a$ to $\tilde M$.
Define $\tilde x^0$ on $\tilde M$ such that
$\nabla_a\tilde x^0 =\mu^{-1}\xi_a-\tau_a$ (note the right
side is curl-free and yields 1 when contracted with 
$\tilde \xi^a$).
Then the hyperbolic coordinates associated with
the $\tilde x^\mu$ coordinates yield a completion of $\tilde M$
in the sense above.

\noindent \hspca$^{13}$ Such freedom in choices of (inequivalent) completion
are known to exist also for other frameworks
such as that of Geroch and that of Ashtekar-Hansen.

\noindent \hspca$^{14}$ P. G. Bergmann, Phys. Rev. {\bf 124}, 274, (1961).

\noindent \hspca$^{15}$ P. T. Chrusciel, J. Math. Phys. {\bf 30}(9), 2094, (1989).

\noindent \hspca$^{16}$ More generally, for a spin-$s$ field, 
we would demand asymptotic
regularity  of order $s+1$.

\noindent \hspca$^{17}$ To see this, evaluate $D_{[a}D_{b]}(
\eta^{\mu\nu} D_c\alpha_\mu D_d\alpha_\nu )$ using eqn.(\ref{tr})
and equate the result to 
$2{\cal R}_{ab(c}{^m} (\eta^{\mu\nu}$
$D_{d)}\alpha_\mu D_m\alpha_\nu )$, 
to obtain
$\eta^{\mu\nu} D_a\alpha_\mu D_b\alpha_\nu =
\eta^{\mu\nu} \alpha_\mu \alpha_\nu \met{q}{0}{_{ab}}$.
Now contract with $\met{q}{0}{^{ab}}$ using 
eqn.(\ref{tr1}). 

\noindent \hspca$^{18}$ One might be tempted to consider, in addition,
those divergence-free vector fields that are 
multi-linear in the Killing fields
on ${\cal H}$.   However, this adds nothing new
since every anti-symmetric second rank tensor
$F^{\mu\nu}$ in ${\cal T}$ yields a Killing field in ${\cal H}$
when contracted with $\alpha_{[\mu}D^a\alpha_{\nu ]}$
and, conversely, for every Killing field $\xi^a$ in ${\cal H}$,
there exists an anti-symmetric second rank tensor over ${\cal T}$ 
(namely
$F_{\mu\nu}
\equiv 2 \xi_a \alpha_{[\mu}D^a\alpha_{\nu ]}
+ D_a\xi_b D^a\alpha_\mu D^b\alpha_\nu$)
that gives rise to it.
Similarly, multi-linearity in conformal Killing fields
yields nothing new, for every vector 
$v^{\mu}$ in ${\cal T}$ yields a 
curl-free conformal Killing field 
in ${\cal H}$
when contracted with $D^a\alpha_\mu $ and conversely,
for every curl-free conformal Killing field $\zeta^a$
in ${\cal H}$, 
there exists a vector over ${\cal T}$ 
(namely
$v^\mu = \zeta^a D_a\alpha^\mu - \frac{1}{3}(D_a\zeta^a)\alpha^\mu$)
that gives rise to it.

\noindent \hspca$^{19}$ In Ref. 10, Ashtekar and Romano used instead
the condition 
$\lim_{\Omega\rightarrow 0} \Omega^{-1}\tilde G_{ab}=0$
to show that the angular momentum is conserved. 
As we have noted earlier, their condition is too strong.
The condition we are imposing is the necessary and sufficient
condition for $\met{B}{1}{_{ab}}$
to be divergence-free
on ${\cal H}$ (c.f. eqn.(\ref{eqb1})).
An example of a space-time satisfying our additional
condition is the Kerr-Newman solution.
In fact, the Kerr-Newman solution satisfies 
a stronger condition: $\met{T}{0}{_a} =0$.
In general, it is not clear how restrictive
is the condition given by eqn.(\ref{angcd}). 
However, the condition is presumably satisfied for
all stationary asymptotically flat space-times since in
that case one expects the angular momentum is well-defined
and equal to Hansen's angular-momentum dipole moment$^{36}$.

\noindent \hspca$^{20}$ To see this, note that 
\begin{eqnarray}
{\cal M}'{_{\mu\nu}}
- 
{\cal M}_{\mu\nu}
&=&
-\frac{1}{16\pi}
\epsilon_{\mu\nu} {^{\tau\sigma }}
\int_C \{
(-2 \epsilon{^{mn(a}}\metbar{E}{0}{^{b)}}{_m}D_n 
\met{\omega}{0}
 )\alpha_{\tau }
D_b\alpha_{\sigma } \}dS_a
\nonumber \\ &=&
-\frac{1}{8\pi}
\met{\omega}{0}{_{[\mu} }
\int_C \metbar{E}{0}{^{ab}}D_{|b|}
\alpha_{\nu ]} dS_a
\nonumber \\ &=&
- \met{\omega}{0}{_{[\mu} }
{\cal P}_{\nu ]} .
\nonumber
\end{eqnarray}

\noindent \hspca$^{21}$ The linearity is clear for electric charge. For
total energy-momentum and angular momentum, we have
in mind the linearized
gravity in Minkowski space-time in which these quantities
are linear in the gravitational field and are expressible
as surface integrals$^{38}$

\noindent \hspca$^{22}$ That is, 
\begin{equation}
{\cal C} (\alpha_{\mu_1} \cdots \alpha_{\mu_{n-1}})
\equiv 
\sum_{m=0}^{[n/2]}
\left (
-\frac{1}{4}
\right )
^m
\left ( \begin{array}{c} n-m-1 \\ m \end{array} \right )
\eta_{({\mu_1}{\mu_2}}
\cdots\eta_{{\mu_{2m-1}}{\mu_{2m}}}
\alpha_{{\mu_{2m+1}}}\cdots
\alpha_{{\mu_{n-1}})},
\nonumber
\end{equation}
with $[n/2]$ denoting the largest integer 
not exceeding $n/2$.

\noindent \hspca$^{23}$ Indeed, we have
\begin{align*}
&D^a (\met{E}{n}{^m}D_m\alpha_{(\mu} )
{\cal C}(\alpha_{\mu_1}
\cdots \alpha_{\mu_{n-1})} )
- \met{E}{n}{^m}D_m\alpha_{(\mu }
D^a  {\cal C}(\alpha_{\mu_1}
\cdots \alpha_{\mu_{n-1})} ) 
\\
=&D_b 
\left \{
2\met{E}{n}{^{[a}}D^{b]}\alpha_{(\mu}
{\cal C}(\alpha_{\mu_1} \cdots \alpha_{\mu_{n-1})} )
+\epsilon^{abc}\met{B}{n}{_c}
\left [\alpha_{(\mu}
{\cal C}(\alpha_{\mu_1} \cdots \alpha_{\mu_{n-1})} )
-\frac{1}{2}\eta_{(\mu\mu_1}
{\cal C}(\alpha_{\mu_2} \cdots \alpha_{\mu_{n-1})} ) 
\right ]
\right \},
\end{align*}
which can be seen by using eqn.(\ref{id1}) and the identity
\begin{eqnarray}
\alpha_{(\mu}D^a
{\cal C}( \alpha_{\mu_1} \cdots \alpha_{\mu_{n-1})})
&=&
(n-1) {\cal C}( \alpha_{(\mu_1} \cdots \alpha_{\mu_{n-1}})
D^a \alpha_{\mu )}
+\frac{1}{2}\eta_{(\mu\mu_1} D^a
{\cal C}( \alpha_{\mu_2} \cdots \alpha_{\mu_{n-1})}).
\nonumber 
\end{eqnarray}

\noindent \hspca$^{24}$ To see this, note that the difference between the integrands
of ${^{*}{\cal E}}_{\mu\mu_1\cdots\mu_{n-1}}$
and that of the right of eqn.(\ref{edual})
is given by
\[ 
D_k 
\left \{
\frac{2}{n}D_m\alpha_\mu \epsilon^{lm[a}
\left [
(D^{k]}\met{E}{n}{_l})
{\cal C}(\alpha_{\mu_1} \cdots \alpha_{\mu_{n-1}} )
- \met{E}{n}{_l}D^{k]}
{\cal C}(\alpha_{\mu_1} \cdots \alpha_{\mu_{n-1}} )
\right ]
\right \}.
\]
Integrate over a cut of ${\cal H}$. 

\noindent \hspca$^{25}$ To see this, note that the difference between the integrands
of 
${^{*}{\cal G}}_{\mu\nu\mu_1\cdots\mu_{n-1}}$
and the right side of eqn.(\ref{gdual})
is given by
\begin{align} 
&D_k 
\left \{ \frac{1}{n}\epsilon^{akl}\met{E}{n}{_{lm}}\alpha_{(\mu}
D^m\alpha_{\nu )} {\cal C}(\alpha_{\mu_1} \cdots \alpha_{\mu_{n-1}} )
\right .
 \nonumber \\
 &\hspace{1.2cm}
\left .
+ \frac{2}{n}
D_m\alpha_{(\mu }
D^{s}\alpha_{\nu )} 
\epsilon^{lm[a}
\left [
(D^{k]}\met{E}{n}{_{ls}})
{\cal C}(\alpha_{\mu_1} \cdots \alpha_{\mu_{n-1}} )
- \met{E}{n}{_{ls}}D^{k]}
{\cal C}(\alpha_{\mu_1} \cdots \alpha_{\mu_{n-1}} )
\right ]
\right \}.
\nonumber 
\end{align} 
Integrate over a cut of ${\cal H}$.

\noindent \hspca$^{26}$ Indeed, we have the following.
Let $\met{\phi}{n-2}$ satisfy
the $n-2$-th remnant equation for a Klein-Gordon field,
$\xi^a$ any Killing field. Denote by $\met{\psi}{n-2}_{a_1\cdots a_s}$
the symmetric and trace-free part 
of $D_{a_1}\cdots D_{a_s}\met{\phi}{n-2}$. Then
\begin{align*}
\met{E}{n}{_{ab}}&\equiv 
\met{\psi}{n-2}_{abcd}\xi^c \xi^{*d}
+\frac{12}{7}(n+1)(n+2)[
\xi^{*m}\met{\psi}{n-2}_{m(a}\xi_{b)} 
+\xi^{m}\met{\psi}{n-2}_{m(a}\xi^*_{b)} 
-\frac{2}{3}\met{q}{0}{_{ab}}
\met{\psi}{n-2}{^{cd}} \xi^{c} \xi^{*d} ] \\
&\ -\frac{4}{5}n(n+1)(n+2)\met{\psi}{n-2}{^m}D_m( \xi_{(a} \xi^{*}_{b)} )
+\frac{4}{5}n(n-1)(n+1)(n+2)\met{\psi}{n-2} \xi_{(a} \xi^{*}_{b)}  
\\
&\ +(n+2)\met{\psi}{n-2}{_{cd(a}}D_{b)}( \xi^{(c} \xi^{*}{^{d)}} )
\end{align*}
satisfies the $n$-th remnant equations for a linearized gravitational field.

\noindent \hspca$^{27}$ Adding to $\psi$ a constant changes 
the integrand of ${\cal E}_{\mu\nu}$ in eqn.(\ref{cemunu})
by a divergence of an antisymmetric tensor.

\noindent \hspca$^{28}$ Let $\tilde\phi$ be a Klein-Gordon field in Minkowski space-time
asymptotically regular of order 1. Consider
\[
I = \int_{S_\infty} - \tilde \epsilon_{abcd} x^c\tilde\nabla^d
[(x^e\tilde\nabla_e + 1)\tilde\phi ],
\]
where $S_\infty$ denotes a two-sphere at infinity, 
and $x^a$ a position vector field.
When $S_\infty$ is any two-sphere cut at spatial infinity, 
the above integral reproduces the remnant radiation
monopole ${\cal K}$ associated with $\tilde\phi$. However,
in general the integral evaluates to a different value
when the two-sphere $S_\infty$ is at null infinity.
For example, let $\tilde\phi =(f(t+r)-f(t-r))/r$, with
$k_\pm (x)\equiv f(\pm \frac{1}{x}), x>0$, both smoothly
extendible to zero. 
Then, when $S_\infty$ is at spatial infinity, we have
$I = {\cal K}=4\pi [k'_+(0) +k'_-(0)]$,
while for $S_\infty$ any cut at future null infinity, 
$I = 4\pi k'_+(0) $, and
for $S_\infty$ any cut at past null infinity, 
$I = 4\pi k'_-(0) $.

\noindent \hspca$^{29}$ Newman and Penrose, Proc. R. Soc. London, A {305}, 175 (1968).

\noindent \hspca$^{30}$ A. R. Exton, E. T. Newman, and R. Penrose, J.
Math. Phys. {\bf 10}, 1566, (1969).

\noindent \hspca$^{31}$ The analogous equation for $\met{B}{n}{_{ab}}$,
\begin{equation}
{\cal L}_{D\zeta } \met{B}{n}_{ab} 
-(n+1)\zeta\met{B}{n}_{ab}
-2\epsilon^{lm}{_{(a}}\met{E}{n}{_{b)l}}D_m\zeta =0 ,
\nonumber
\end{equation}
follows from eqn.(\ref{grstatic}) and the remnant field equations.
For a Maxwell field the corresponding equation is 
\begin{equation}
{\cal L}_{D\zeta} \met{E}{n}_{a} 
-(n+1)\zeta\met{E}{n}_{a}
+\epsilon_a{^{kl}}\met{B}{n}{_{k}}D_l\zeta =0 .
\nonumber
\end{equation}
and for a Klein-Gordon field,  eqn.(\ref{static}).

\noindent \hspca$^{32}$ We remark that the multipole moment $M$ defined here is
related to the $Q$ defined by Geroch in Ref. 38
by a normalization factor: 
$Q_{i_1\cdots i_n}=(-\frac{1}{3})^nn!M_{i_1\cdots i_n}$.

\noindent \hspca$^{33}$ We are concerned only with ``irreducible'' solutions.
Thus, for example,
${\cal C}( \alpha_{\mu_1}\cdots \alpha_{\mu_{n-1}})$
and 
$\eta_{\mu\nu} {\cal C}( \alpha_{\mu_1}\cdots \alpha_{\mu_{n-1}})$
are viewed as equivalent solutions to
eqn.(\ref{phin}).

\noindent \hspca$^{34}$ To see this, embed ${\cal H}$ as the unit hyperboloid in Minkowski
space-time $M'$. Let $x^a$ denote the position vector field from 
some origin.
Then $\nabla_a x^b =\delta_a{^b}$ and ${\cal H}$ is specified by
$x^ax_a =1$.
Let $k^a$ be a constant vector field in $M'$.
Then $k^ax_a$ is a translation on ${\cal H}$.
Thus the most general function multi-linear in translations is
a sum of terms of the form $w(s)\equiv w^{a_1\cdots a_s}
x_{a_1} \cdots x_{a_s}$, with $w_{a_1\cdots a_s}$ some symmetric,
trace-free constant tensor. This $w(s)$ 
satisfies the Klein-Gordon equation in $M'$.
Using $\nabla^2 w= [D^2 +(x\cdot x)^{-1}(
(x\cdot\nabla )^2
+2x\cdot\nabla )]w$, we see that $w(n-1)$ satisfies eqn.(\ref{phin})
on ${\cal H}$.
Such $w(n-1)$'s clearly exhaust all solutions of
eqn.(\ref{phin}) which are multi-linear in translations.
But each such $w(n-1)$'s on ${\cal H}$
is the contraction of 
${\cal C}( \alpha_{\mu_1}\cdots \alpha_{\mu_{n-1}})$
with some tensor over ${\cal T}$.

\noindent \hspca$^{35}$ One way to prove the identity is to note:
$D^a\alpha_\mu D_a
{\cal C}( \alpha_{\mu_1} \cdots \alpha_{\mu_{n-1}})
+(n-1)
\alpha_\mu {\cal C}( \alpha_{\mu_1} \cdots \alpha_{\mu_{n-1}})$
satisfies eqn.(\ref{phin}) with $n$ replaced by $n-1$ and
is trace-free in $\mu_1\cdots\mu_{n-1}$.
The overall normalization factor can be fixed by comparing, say,
the coefficients of the term $\eta_{(\mu \mu_1}
{\cal C}( \alpha_{\mu_2} \cdots \alpha_{\mu_{n-1})})$ on both sides.

\noindent \hspca$^{36}$ R. O. Hansen, J. Math. Phys. {\bf 15}, 46, (1974).

\noindent \hspca$^{37}$ R. Geroch, J. Math. Phys. {\bf 12}, 918, (1970).

\noindent \hspca$^{38}$ R. Penrose, Proc. R. Soc. London, A {\bf 381}, 53, (1982).

\noindent \hspca$^{39}$ R. Geroch, J. Math. Phys. {\bf 11}, 1955, (1970).

\noindent \hspca$^{40}$ A. Ashtekar and A. Magnon-Ashtekar, Phys. Rev. Lett.
{\bf 43}(3), 181 (1979).

\noindent \hspca$^{41}$ R. Beig and W. Simon, Comm. Math. Phys. {\bf 78}, 75, (1980).

\noindent \hspca$^{42}$ P. G. Bergmann, Gen. Rel. Grav., {\bf 19}, 371, (1987).

\noindent \hspca$^{43}$ P. G. Bergmann, Gen. Rel. Grav., {\bf 21}, 271, (1988).

\noindent \hspca$^{44}$ P. G. Bergmann, Phys. Rev. D {\bf 48}, 5684, (1993).

\noindent \hspca$^{45}$ R. Geroch, {\em Asymptotic structure of space-time},
    ed. P. Esposito and L. Witten (New York, Plenum, 1976).

\noindent \hspca$^{46}$ J. N. Goldberg, Phys. Rev. D {\bf 41}, 410, (1990).

\noindent \hspca$^{47}$ C. Hoenselaers, Prog. Theor. Phys. {\bf 55}, 406, (1976).

\noindent \hspca$^{48}$ R. Penrose, Proc. R. Soc. London, A {\bf 284}, 159, (1965).

\noindent \hspca$^{49}$ R. K. Sachs, Proc. R. Soc. London, A {\bf 270}, 193, (1962).

\noindent \hspca$^{50}$ J. Winicour, Found. Phys. {\bf 15}, 605, (1985).

\end{document}